\documentclass[11pt]{article}
\usepackage{epsfig}
\usepackage{amscd,amsfonts,amsmath,latexsym, amssymb, amsthm}
\usepackage{natbib}
\usepackage{eucal}

\theoremstyle{plain}
\newtheorem{theorem}{Theorem}[section]
\newtheorem{proposition}[theorem]{Proposition}

\theoremstyle{plain}

\theoremstyle{plain}
\newtheorem{remark}{Remark}[section]

\theoremstyle{definition}

\theoremstyle{definition}

\theoremstyle{definition}

\title{Maximising Survival, Growth, and Goal Reaching Under Borrowing Constraints}
\author{Haluk Yener \thanks{Email. haluk.yener@bilgi.edu.tr }\\Istanbul Bilgi University \\Dept. of Business Administration, Santral Campus \\ 34060, Eyup, Istanbul, TURKEY}
\date{ This version: November 19, 2011}

\begin{document}
\maketitle

\\

\noindent\textit{MSCcodes}: Primary: 93E20, 60H10, ; Secondary: 60G40, 60J60. 
\\

\noindent\textit{ORMScodes}: Primary: Probability, Dynamic Programming/Optimal Control; Secondary: Finance.

\begin{abstract}
In this paper, we consider three problems related to survival, growth, and goal reaching maximization of an investment portfolio with proportional net cash flow. We solve the problems in a market constrained due to borrowing prohibition. To solve the problems, we first construct an auxiliary market and then apply the dynamic programming approach. Via our solutions, an alternative approach is introduced in order to solve the problems defined under an auxiliary market. 
\end{abstract}

\section{Introduction.} \label{intro} The paper involves the application of stochastic modelling to portfolio optimization in an auxiliary market. Portfolio optimization is related to finding optimal investment strategies within a portfolio composed of dynamically traded risky assets. The optimality of these strategies depend on the specification of an objective function that is related to the aim of an investor and to any possible constraints (i.e. investment, borrowing, and short-selling constraints) that may exist in the market. The objective function, in turn, is generally given by a subjective utility function and involves either the maximization or the minimization of it. By the application of the dynamic programming or martingale approach, it is then possible to find the optimal results.

In the continuous time finance literature, the dynamic programming approach to a general class of utility functions was first applied by the pioneering work of \citet{Merton}. The application of martingale approach, on the other hand, was introduced by \citet{Pliska}. Extensions to both approaches involve vast amount of studies. Notable ones are the work of \citet{DavisNorman}, \citet{Zar1}, \citet{ShreveSoner}, and \citet{CviKar} on optimal consumption and portfolio selection under transaction costs. The works of \citet{FlemingZar}, and \citet{Zar2} involve optimal portfolio selection when borrowing rate is higher than the lending rate and trading constraints respectively. Constraints on trading are also considered by the works of \citet{ShreveXu}, \citet{CvitanicKaratzas}, and \citet{KaratzasKou}. Furthermore,\citet{CviCuo} studies the optimal consumption problem of a large investor whose actions affect the market. See \citet{KaratzasShreve} and \citet{Cvitanic} for more details on the extensions.
 
In this paper, we add to the extensions by considering borrowing constraints for problems concerning survival, growth, and goal reaching maximization in terms of probability, time, and expected discounted reward. A previous application of these problems was done by \citet{Browne3} to find the optimal investment strategies for outperforming a benchmark strategy. He considered infinite time approach and the market is in a sense incomplete because the benchmark satisfies a stochastic process that is not perfectly correlated with the investment opportunities. Under this scenario, he first solves the probability maximization problem for beating a benchmark. After solving this problem, Browne finds the optimal investment strategy that minimises the expected time to beat a benchmark. He also solves the time maximization problem for staying above the lower boundary. Finally, Browne considers maximising or minimising the expected discounted reward. The maximization is relevant when there is a reward for achieving a goal and minimization is relevant when there is a penalty to pay. 

In the current paper, the analysis of \citet{Browne3} is applied without considering any benchmark process. In this form, the approach is a different version of the work by \citet{Browne2} that solves the optimal investment strategy of an investor who is constantly withdrawing money from an account. In that paper, Browne divides the investment region into two parts via creating a boundary level from the perpetual value of the cash withdrawals. If the wealth level is below the perpetual value, then, the investor is said to be in the {\em danger} zone because initial cash value is not large enough to cover the perpetual withdrawals. As a result, she may go into bankruptcy. Under this scenario, the goal of the investor is to find the optimal investment strategy that maximises her survival. Browne shows that when in the danger zone there is no optimal investment that eliminates the possibility of bankruptcy. On the other hand, if the wealth level is above the perpetual cash withdrawals, the investor is said to be in the {\em safe} zone. Then, bankruptcy will be avoided with certainty, and the investor will aim to find the strategy that maximises the growth for reaching a goal. 

The case considered in this paper differs from the one considered above, because it applies a {\em constant proportional} net cash flow rate. In addition, the wealth process is formed under the constrained market scenario. As mentioned previously constraints happen due to borrowing prohibition. This is the case analysed by \citet{BayraktarYoung} for the minimization of lifetime ruin. First, they consider the restriction of borrowing and short selling. Then, they proceed to the problem where the investor is allowed to borrow money only at a rate that is higher than the rate earned on the riskless rate. To this end, they consider, as in \citet{Young} (which only considers minimising the probability of lifetime ruin), constant consumption and constant proportional consumption rate. The wealth portfolio consists of investment in a risky and a riskless asset. They find the solution by rescaling the objective function. This work, on the other hand, involves investment into multiple risky assets. In addition, an auxiliary market is constructed and a dynamic programming approach is used in order to solve the problems. Details on the use of an auxiliary market can also be seen in \citet{KaratzasShreve} and \citet{Cvitanic}. To solve the problems they consider a stochastic duality argument under the martingale approach. Our approach, however, contributes to the literature by applying the dynamic programming method to three problems related to survival, growth, and goal reaching.

In order to show the results, the assets and the model are introduced in section \ref{model3} first. Then, the problems are solved. The first results are provided in section \ref{maximizeprobability3}. There, the optimal value function and the investment strategy for reaching a goal without first hitting a lower barrier is found. Next, the solutions for maximising the time to survive and minimising the time to achieve a goal are provided in section \ref{3problem2} . Finally, the solutions to the problem of maximising and minimising the expected discounted reward are provided in section \ref{3problem3}. In all problems, the verifications are done by using a variation of the method in \citet{Bjork}.

\section{The Assets and The Model.} \label{model3} In this section we introduce the market model along with the processes and the financial assets that we use to create an investor's wealth process. We first give the background on the market model, then, we complete the markets by creating an auxiliary market with fictitious assets. 

We consider infinite time horizon and assume that the market is modelled under a filtered probability space $(\Omega,\mathcal{F},\{\mathcal{F}_{t}\}_{0 \leq t < \infty}, \mathbb{P})$. The market filtration is spanned by a $N$-dimensional standard Brownian motion $B(t):=\left(B_{1}(t), \ldots, B_{N}(t); t < \infty \right)$ which is defined on our complete probability space $(\Omega,\mathcal{F},\{\mathcal{F}_{t}\}_{0 \leq t < \infty}, \mathbb{P})$. Here, $\mathbb{P}$ is the real measure and $\{\mathcal{F}_{t}\}_{0 \leq t < \infty}$ is the $\mathbb{P}$-augmentation of the natural filtration $\mathcal{F}_{t}^{B}:=\sigma \{B(u) \mid u \leq t\}$.  

On the other hand, we assume that there is an investor who trades continuously in a Black-Scholes type frictionless financial market. The traded assets are risky stocks and a riskless asset. We denote the riskless asset by creating a bank account process of the form
\begin{equation}
dV_{0}(t)=rV_{0}(t)dt, \label{3eq1}
\end{equation}

\noindent where the riskless rate $r$ is a constant. On the other hand, the stocks are given by
\begin{equation}
dS_{i}(t)=S_{i}(t)\left[\mu_{i}dt + \sum_{j=1}^{N}\sigma_{ij}dB_{j}(t)\right] \quad \mbox{for} \quad i=1, \ldots, N, \label{3eq2} 
\end{equation}

\noindent where $\mu_{i}$ and $\sigma_{ij}$, for $i,j=1, \ldots, N$, are constants. Moreover, we assume that there is a net cash flow process which is {\em proportional} to the wealth process. That is, the net cash flow amount at time $t$ is given by $C^{net}(t,X^{w}(t))=c^{net}X^{w}(t)$, where $c^{net}$ is a {\em negative} constant that denotes the net cash flow rate, and $X^{w}(t)$ is the wealth process value at time $t$. 

We express the wealth process by using a vector of control processes $\textit{\textbf{w}}(t):=\left(w_{1}(t), \ldots, w_{N}(t) \right)^{'}$ that represents {\em the proportions of wealth} invested in the risky assets at time $t$. We call $\textit{\textbf{w}}(t)$ an investment strategy and say that the strategy is admissible for a initial capital amount $x$, that is $\textit{\textbf{w}} \in \mathcal{A}(x)$, if $\textit{\textbf{w}}(t)$ is $\{\mathcal{F}_{t}\}$-progressively measurable, satisfies $\int_{0}^{t}\|\textit{\textbf{w}}(s)\|^{2}ds < \infty$ almost surely for $t < \infty$. Therefore, the {\em self-financing} wealth process associated to an admissible strategy is the solution of the stochastic differential equation
\begin{eqnarray}
dX^{w}(t)&=&X^{w}(t)\left[\left(r+c^{net}\right)dt+\textit{\textbf{w}}^{'}(t)(\mu - r1_{N})dt+\textit{\textbf{w}}^{'}(t)\sigma dB(t)\right]; \nonumber \\
X(0)&=&x \label{3eq3} 
\end{eqnarray}

\noindent with $1_{N}:=\left(1,\ldots,1\right)^{'}$, $\mu:=\left(\mu_{1},\ldots,\mu_{N}\right)^{'}$, $\sigma:=\left(\sigma_{1},\ldots,\sigma_{N}\right)$ and $\sigma_{i}:=\left(\sigma_{i1}, \ldots, \sigma_{iN}\right)^{'}$ for $i=1,\ldots,N$. 

In order to solve the problem when borrowing is prohibited, we create an auxiliary market as outlined in \citet{KaratzasShreve}. We let $\mathcal{K} \neq \emptyset$, $\mathcal{K} \in \mathbb{R}^{N}$ be a closed convex set in which the proportional investment strategies are constrained.

We denote by $\mathcal{A}_{c}(x)$ the set of admissible strategies of the constrained market, and define for a given $\mathcal{K}$, the {\em support function} of the convex set $-\mathcal{K}$ by 
\[\delta(\nu)=\sup_{\textit{\textbf{w}} \in \mathcal{K}}(-\textit{\textbf{w}}^{'}\nu), \quad \nu \in \mathbb{R}^{N}.\]

The support function is finite on its effective domain
\[\tilde{\mathcal{K}}:=\{\nu \in \mathbb{R}^{N} \mid \delta(\nu) < \infty \},\]

\noindent which is also the corresponding barrier cone of $-\mathcal{K}$. We assume that
$\tilde{\mathcal{K}}$ contains the origin on $\mathbb{R}^{N}$, and set $\delta(\nu) \geq 0$ $\forall \nu \in \mathbb{R}^{N}$ with $\delta(0)=0$. We also have $\delta(\nu)+\textit{\textbf{w}}^{'}\nu \geq 0$, $\forall \nu \in \tilde{\mathcal{K}}$ if and only if $\textit{\textbf{w}} \in \mathcal{K}$.

Because we are considering borrowing prohibition as the constraint, we let $\delta(\nu)=-\nu_{1}$ on $\tilde{\mathcal{K}}$ for some scalar $\nu_{1} \leq 0$, and define the constraint set by
\[\mathcal{K}:=\{\textit{\textbf{w}} \in \mathbb{R}^{N} \mid \sum_{i=1}^{N}w_{i} \leq 1 \}. \]

\noindent Then, the corresponding barrier cone is
\[\tilde{\mathcal{K}}:=\{\nu \in \mathbb{R}^{N} \mid \nu_{1}=\ldots=\nu_{N} \leq 0 \}.\]

Next, we express the assets of the auxiliary market by 
\begin{eqnarray}
dV_{0}(t)&=&V_{0}(t)\left(r+\delta(\nu)\right)dt; \label{3eq5} \\
dS_{i}^{\nu}(t)&=&S_{i}^{\nu}(t)\left[\left(\mu_{i}+\nu_{i}+\delta(\nu)\right)dt+\sum_{j=1}^{N}\sigma_{ij}dB_{j}(t)\right]  \qquad \mbox{for} \quad i=1, \ldots, N. \label{3eq6}
\end{eqnarray}

\noindent In this market, the wealth process is the solution of the stochastic differential equation for $\nu \in \tilde{\mathcal{K}}$
\begin{equation}
dX^{w}_{\nu}(t)=X^{w}_{\nu}(t)\left[\left(r+\delta(\nu)+c^{net}\right)dt+\textit{\textbf{w}}^{'}(t)\left(\mu + \nu - r1_{N}\right)dt+\textit{\textbf{w}}^{'}(t)\sigma dB(t)\right]. \label{3eq7}
\end{equation}

On the other hand, the market price of risk under the fictitious market is
\begin{eqnarray}
\zeta_{\nu}&=&\sigma^{-1}(\mu+\nu-r1_{N}) \nonumber \\
&=& \zeta+\sigma^{-1}\nu, \qquad \qquad \nu \in \tilde{\mathcal{K}}, \label{3eq8}
\end{eqnarray}

\noindent where $\zeta=\sigma^{-1}(\mu-r1_{N})$ is the market price of risk under the constrained market.

Finally, the generator of $X^{w}_{\nu}(\cdot)$ for every open set $\mathcal{O} \in \mathbb{R}$, for functions $\Gamma_{\nu}(x) \in C^{2}(\mathcal{O})$, and for every control process $\textit{\textbf{w}} \in \mathbb{R}^{N}$ is given for $\nu \in \tilde{\mathcal{K}}$ by
\begin{equation}
\mathcal{L}^{w}\Gamma_{\nu}(x)=\left(\left(r+\delta(\nu)+c^{net}\right)+\textit{\textbf{w}}^{'}(\mu +\nu - r1_{N})\right)x\frac{\partial}{\partial x}\Gamma_{\nu}(x)+\frac{1}{2}\textit{\textbf{w}}^{'}\Sigma\textit{\textbf{w}}x^{2}\frac{\partial^{2}}{\partial x^{2}}\Gamma_{\nu}(x).
\label{3eq9}
\end{equation}

\noindent where $\Sigma=\sigma\sigma^{'}$. We also let $\mathbb{E}_{x}\left[\cdot \right]=\mathbb{E}\left[\cdot \mid X(0)=x\right]$ throughout the text. 

\section{Maximising The Probability of Hitting a Target Before Default.} \label{maximizeprobability3} In this subsection, we provide the formulation for the optimal value and the optimal strategy for reaching a target level before hitting a default boundary. To do this, we start by expressing for $\nu \in \tilde{\mathcal{K}}$ the drift and the volatility of the equation (\ref{3eq7}) by
\begin{eqnarray}
m(x,\textit{\textbf{w}}(x))&=&\left(r+\delta(\nu)+c^{net}\right)x+\textit{\textbf{w}}^{'}(x)\left(\mu +\nu - r1_{N}\right)x; \nonumber \\
v(x,\textit{\textbf{w}}(x))&=&\textit{\textbf{w}}^{'}(x)\Sigma\textit{\textbf{w}}(x)x^{2}. \label{3eq10}
\end{eqnarray} 

\noindent We see that the wealth can diminish towards undesired levels due to the presence of a negative net cash flow rate $c^{net}$. Especially, when $r+\delta(\nu)+c^{net}$ is negative the investor has to take risk so that her wealth can be kept from falling into undesired levels. If the investor doesn't take any risk, her wealth process has the form
\[X_{\nu}(t)=xe^{(r+\delta(\nu)+c^{net})t}.\]

\noindent However, when the term $r+\delta(\nu)+c^{net}$ is negative, $X_{\nu}(t)$ eventually approaches zero as the time progresses.

In addition, if there is a target greater than $x$, it can only be reached without taking risk if the term $r+\delta(\nu)+c^{net}$ is positive. However, with $r+\delta(\nu)+c^{net} < 0$, reaching a target by staying above an undesired level cannot solely be realised by simply investing in the bank account. In this case, our investor must take risk and look for strategies that will increase her chances to hit an upper barrier level without first hitting a lower barrier level. In other words, if the investor manages to reach a target level $U$ before hitting the default barrier $L$, for $0 < L < x < U$, she will realize her objective. To model the objective, we let
\[\tau_{U}^{w}=\inf\{t > 0 \mid X^{w}_{\nu}(t) \geq U\}\]

\noindent be the first time the portfolio process crosses the upper barrier, and
\[\tau_{L}^{w}=\inf\{t > 0 \mid X^{w}_{\nu}(t) \leq L\}\]

\noindent be the first time the portfolio process crosses the lower barrier. 

\begin{remark}
By using the log-optimal strategy $\textbf{w}_{o}=\sigma^{-1}\zeta_{\nu}$ for $\nu \in \tilde{\mathcal{K}}$ in the auxiliary market, the wealth process of an investor can be expressed by
\begin{equation}
X^{w_{o}}_{\nu}(t)=x\exp\left\lbrace \beta_{\nu} t +\zeta_{\nu} B(t) \right\rbrace \label{3remark1}
\end{equation}

\noindent where 
\begin{equation}
\beta_{\nu}=(r+c^{net}+\delta(\nu))+\frac{1}{2}\|\zeta_{\nu}\|^{2}. \label{3remark2}
\end{equation}

\noindent We name $\beta_{\nu}$ as the market favourability parameter. When $\beta_{\nu} < 0$, we say that the markets are unfavourable and when $\beta_{\nu} > 0$ we say that the markets are favourable. Hitting the lower barrier level $L$ as the time passes is more likely under the unfavourable markets than it is under the favourable markets. That is, the time is to the disadvantage of the investor in unfavourable markets. To see it more clearly, we write
\begin{equation}
\mathbb{P}\left(X^{w_{o}}_{\nu}(t) \leq L \right) = \Phi \left(\frac{\ln\left(\frac{L}{x}\right)-\beta_{\nu} t}{\sqrt{\|\zeta_{\nu}\|^{2} t}}\right) \label{3remark3}
\end{equation}

\noindent where $\Phi(.)$ is the normal cdf. We see that $\mathbb{P}\left(X^{w_{o}}_{\nu}(t) \leq L \right) \rightarrow 1$ as $t \rightarrow \infty$ when $\beta_{\nu} < 0$. On the other hand, when the markets are favourable the situation is reversed and the time is to the advantage of the investor.
\end{remark}

As we see from the above remark, in an unfavourable market, we would expect the investor to try harder in order to achieve her objective since the time is to her disadvantage. In any case, she will always seek for strategies which will increase the probability that the first barrier reached is the upper barrier. Therefore, the objective is given by
\begin{equation}
F_{\nu}(x)=\sup_{\textit{\textbf{w}} \in \mathcal{A}_{\nu}}\mathbb{P}_{x}\left(\tau_{U}^{w} < \tau_{L}^{w}\right),\label{3pro1}
\end{equation}

\noindent where $\mathbb{P}_{x}\left(\cdot \right)=\mathbb{P}\left(\cdot \mid X(0)=x\right)$ and $\mathcal{A}_{\nu}(x)$ is the set of admissible strategies in the {\em auxiliary} market. 

\begin{remark}
The goal of the investor is to find a strategy that maximises the probability of hitting $U$ without first hitting $L$. From the way we specified the problem, we had both stopping times finite since the wealth process is expected to hit a barrier eventually. Log-optimal strategy maximises the growth rate of the portfolio process. Another strategy will cause the growth rate to decline. However, it might change the volatility of the portfolio in a way that leads to an increase in the objective. But, this change will not make both stopping times infinite, since the market conditions will dominate as the time goes on. 
\end{remark}

\begin{remark}
The specifications we provide hold for $\nu \in \tilde{\mathcal{K}}$. However, we need to find the optimal fictitious parameter $\nu^{*}$ so that the results of the constrained market can be specified as well . As shown in the appendix, the specification for $\nu^{*}$ is the one that minimises the value function of the auxiliary market over all $\nu \in \tilde{\mathcal{K}}$. That is, we have
\[F_{c}(x)=F_{\nu^{*}}(x)=\inf_{\nu \in \tilde{\mathcal{K}}}F_{\nu}(x), \]

\noindent where $F_{c}(x)$ is the optimal value function in the constrained market, and is given by
\begin{equation}
F_{c}(x)=\sup_{\textit{\textbf{w}} \in \mathcal{A}_{c}}\mathbb{P}_{x}\left(\tau_{U}^{w} < \tau_{L}^{w}\right)\label{contrainedproblem}
\end{equation}

\noindent Furthermore, as mentioned in the appendix, the above argument can also be utilised in other problems considered in this paper. 
\end{remark}

\begin{proposition} \label{prop1}
Let the wealth process $\{X^{w}_{\nu}(t), 0 \leq t < \infty\}$ be the solution to the stochastic differential equation given in (\ref{3eq7}). Let also $\nu_{1}^{*}$ be the minimizer of the term
\[\frac{1}{2\alpha}\|\zeta+\nu_{1}\sigma^{-1}1_{N}\|^{2}+\nu_{1},\]

\noindent where $\alpha \in (-\infty, 0) \setminus \{-1\}$ is\footnote{Notice that we state the cases by using the value of the optimal investment strategy given by the first line in (\ref{thr4}). That is, $\alpha$ in $-\frac{1}{\alpha}1^{'}_{N}\textbf{w}_{o} \geq 1$ is given by the first line of the specification (\ref{thr3}).}
\begin{equation}
\alpha=\begin{cases} \frac{\|\zeta\|^{2}}{2\left(r+c^{net}\right)} & \mbox{if}\,\, -\frac{1}{\alpha}1^{'}_{N}\textbf{w}_{o} < 1; \\ -2\left(K(r+c^{net})+D\right) & \mbox{if}\,\, -\frac{1}{\alpha}1^{'}_{N}\textbf{w}_{o} \geq 1. \label{thr3} \end{cases}
\end{equation} 

\noindent Then, the optimal fictitious parameter is
\begin{equation}
\nu^{*}=\begin{cases} 0_{N} & \mbox{if}\,\, -\frac{1}{\alpha}1^{'}_{N}\textbf{w}_{o} < 1; \\ -\frac{\alpha + D}{K}1_{N} & \mbox{if}\,\, -\frac{1}{\alpha}1^{'}_{N}\textbf{w}_{o} \geq 1, \label{thr1} \end{cases}
\end{equation} 

\noindent where $0_{N}$ is the $N$-dimensional vector of zeros, $D=\zeta^{'}\sigma^{-1}1_{N}$, $K=1_{N}^{'}\Sigma^{-1}1_{N}$, and the optimal value function is given by
\begin{equation}
F_{\nu^{*}}(x)=\frac{L^{1+\alpha}-x^{1+\alpha}}{L^{1+\alpha}-U^{1+\alpha}} \quad \mbox{for} \,\, x \in \left[L,U\right]. \label{thr2}
\end{equation}

\noindent Then, the optimal investment strategy for $L < x < U$ is equal to
\begin{equation}
\textbf{w}^{*}(x)=\begin{cases}-\frac{1}{\alpha}\textbf{w}_{o} & \mbox{if}\,\, -\frac{1}{\alpha}1^{'}_{N}\textbf{w}_{o} < 1; \\ -\frac{1}{\alpha}(\sigma^{'})^{-1}\left(\zeta - \frac{\alpha + D}{K}\sigma^{-1}1_{N}\right) & \mbox{if}\,\, -\frac{1}{\alpha}1^{'}_{N}\textbf{w}_{o} \geq 1, \label{thr4} \end{cases}
\end{equation} 

\noindent where $\textbf{w}_{o}=\Sigma^{-1}(\mu-r1_{N})$. Given the optimal investment strategy, the optimal wealth process for $t < \tau_{L}^{w^{*}} \wedge \tau_{U}^{w^{*}}$ is
\begin{equation}
X_{\nu^{*}}^{*}(t)=\begin{cases} X(0)\exp\left\lbrace \left[(r+c^{net})\left(-\frac{1}{\alpha}-1\right)\right]t-\frac{1}{\alpha}\zeta^{'} B(t)\right\rbrace & \mbox{if}\,\, -\frac{1}{\alpha}1^{'}_{N}\textbf{w}_{o} < 1; \\ X(0)\exp\left\lbrace \left[c^{net}+r+\frac{D}{K}-\frac{1}{2K}\right]t-\frac{1}{\alpha}\zeta^{'}_{\nu^{*}}B(t) \right\rbrace & \mbox{if}\,\, -\frac{1}{\alpha}1^{'}_{N}\textbf{w}_{o} \geq 1 \label{thr5} \end{cases} 
\end{equation}

\noindent with $\zeta_{\nu^{*}}=\zeta -\frac{\alpha + D}{K}\sigma^{-1}1_{N}$.
\end{proposition}

From the results we see that when $\alpha = -1$, the optimal strategy becomes log-optimal strategy. For $\alpha > -1$ $(<-1)$ the investor takes more (less) risk than the log-optimal strategy. On the other hand, we can see that the optimal investment strategy is a constant proportional strategy that is independent of the current wealth level and the barriers $L$ and $U$. However, it is dependent on all other parameters of the model. Furthermore, the favourable case in the first problem happens when $\alpha<-1$, while the unfavourable case happens when $-1<\alpha <0$. Thus, under the unfavourable condition, the investor takes more risk to maximise the probability of hitting a target before default. 

\begin{proof} We assume that $F_{\nu}: \left[L,U\right] \rightarrow \left[0, 1 \right]$ and $F_{\nu}(x)$ is $C^{2}((L,U))$ with $\frac{\partial} {\partial x}F_{\nu}>0$, $\frac{\partial^{2}} {\partial x^{2}}F_{\nu}<0$\footnote{The arguments of the value functions are hidden in the rest of the paper when necessary to simplify the notation.}. Then, the HJB equation over all strategies is for $\nu \in \tilde{\mathcal{K}}$
\begin{equation}
\left(r+c^{net}+\delta(\nu)\right)x\frac{\partial}{\partial x}F_{\nu}+\sup_{\textit{\textbf{w}}}\left\lbrace \textit{\textbf{w}}^{'}(\mu +\nu- r1_{N})x\frac{\partial}{\partial x}F_{\nu}+\frac{1}{2}\textit{\textbf{w}}^{'}\Sigma\textit{\textbf{w}}x^{2}{\frac{\partial^{2}}{\partial x^{2}}F_{\nu}}\right\rbrace =0 \label{3pro12}
\end{equation}

\noindent subject to the boundary conditions $F_{\nu}(U)=1$ and $F_{\nu}(L)=0$. From (\ref{3pro12}), the maximizer $\textit{\textbf{w}}^{*}$ can be specified as
\begin{equation}
\textit{\textbf{w}}^{*}(x)=-(\sigma^{'})^{-1}\zeta_{\nu}\frac{\frac{\partial}{\partial x}F_{\nu}}{x\frac{\partial^{2}}{\partial x^{2}}F_{\nu}}.\label{3pro14}
\end{equation}

\noindent Substituting (\ref{3pro14}) into (\ref{3pro12}) gives the non-linear partial differential equation
\begin{equation}
\left(r+c^{net}+\delta(\nu)\right)x\frac{\partial}{\partial x}F_{\nu}-\frac{1}{2}\|\zeta+\sigma^{-1}\nu\|^{2}\frac{\left(\frac{\partial}{\partial x}F_{\nu}\right)^{2}}{\frac{\partial^{2}}{\partial x^{2}}F_{\nu}} =0. \label{3pro15}
\end{equation}

\noindent Now, we guess a solution of the form $A_{1}-A_{2}x^{1+\alpha}$ with constants $A_{1}$ and $A_{2}$. By using the boundary conditions, its specification can be given more explicitly by
\begin{equation}
F_{\nu}(x)=\frac{L^{1+\alpha}-x^{1+\alpha}}{L^{1+\alpha}-U^{1+\alpha}} \quad \mbox{for} \,\, x \in \left[L,U\right]. \label{3pro16}
\end{equation} 

\noindent $F_{\nu}(x)$ is an increasing function of the wealth process $X^{w}_{\nu}(\cdot)$. Therefore, $\forall \nu \in \tilde{\mathcal{K}}$, the inequality $F_{\nu^{*}}(x) \leq F_{\nu}(x)$ holds. We proceed by substituting (\ref{3pro16}) into (\ref{3pro15}), and obtain 
\begin{equation}
-(1+\alpha)\frac{x^{1+\alpha}}{L^{1+\alpha}-U^{1+\alpha}}\left[\left(r + c^{net}\right)-\frac{1}{2\alpha}\|\zeta+\sigma^{-1}\nu\|^{2}+\delta(\nu)\right]=0. \label{3pro17}
\end{equation}

\noindent The optimal fictitious parameter is $\nu^{*}=\nu^{*}_{1}1_{N}$ with $\nu^{*}_{1}$ minimising the term
\begin{equation}
\frac{1}{2\alpha}\|\zeta+\nu_{1}\sigma^{-1}1_{N}\|^{2}+\nu_{1}. \label{3pro18}
\end{equation}

\noindent From the above term we can find the values of $\nu_{1}^{*}$ as
\begin{equation}
\nu^{*}_{1}=\begin{cases} 0 & \mbox{if}\,\, 1^{'}_{N}\textit{\textbf{w}}^{*}(x) < 1; \\ -\frac{\alpha + D}{K} & \mbox{if}\,\, 1^{'}_{N}\textit{\textbf{w}}^{*}(x) \geq 1, \label{3pro19} \end{cases}
\end{equation} 

\noindent where $D=\zeta^{'}\sigma^{-1}1_{N}$ and $K=1^{'}_{N}\Sigma 1_{N}$. By using the term in the second case of (\ref{3pro19}), we obtain
\[-\frac{1}{2\alpha}\|\zeta+\nu_{1}^{*}\sigma^{-1}1_{N}\|^{2}=-\frac{\alpha}{2K}.\]

\noindent Then, the values of $\alpha$ are
\begin{equation}
\alpha=\begin{cases} \frac{\|\zeta\|^{2}}{2\left(r+c^{net}\right)} & \mbox{if}\,\, 1^{'}_{N}\textit{\textbf{w}}^{*}(x) < 1; \\ -2\left(K(r+c^{net})+D\right) & \mbox{if}\,\, 1^{'}_{N}\textit{\textbf{w}}^{*}(x) \geq 1, \label{3pro20} \end{cases}
\end{equation}

\noindent and the maximizer is
\begin{equation}
\textit{\textbf{w}}^{*}(x)=\begin{cases}-\frac{1}{\alpha}\textit{\textbf{w}}_{o} & \mbox{if}\,\, -\frac{1}{\alpha}1^{'}_{N}\textit{\textbf{w}}_{o} < 1; \\ -\frac{1}{\alpha}(\sigma^{'})^{-1}\left(\zeta - \frac{\alpha + D}{K}\sigma^{-1}1_{N}\right) & \mbox{if}\,\, -\frac{1}{\alpha}1^{'}_{N}\textit{\textbf{w}}_{o} \geq 1. \label{3pro21} \end{cases}
\end{equation} 

\noindent Now, we verify that (\ref{3pro16}) and (\ref{3pro21}) are optimal. First, we check if $\textit{\textbf{w}}^{*} \in \mathcal{K}$. We have when $-\frac{1}{\alpha}1^{'}_{N}\textit{\textbf{w}}_{o} \geq 1$
\begin{eqnarray*}
1^{'}_{N}\textit{\textbf{w}}^{*}(x)&=&-\frac{1}{\alpha}1^{'}_{N}(\sigma^{'})^{-1}\left(\zeta - \frac{\alpha + D}{K}\sigma^{-1}1_{N}\right) \\
&=&-\frac{1}{\alpha}\left(D-\alpha + D\right)=1.
\end{eqnarray*}

\noindent In addition, when $-\frac{1}{\alpha}1^{'}_{N}\textit{\textbf{w}}_{o} < 1$, we have $\nu^{*}=0$ giving us the first term in (\ref{3pro21}). Therefore, $\textit{\textbf{w}}^{*} \in \mathcal{K}$. For such $\textit{\textbf{w}}^{*}$, we have
\[-\nu^{*}_{1}+1_{N}^{'}\textit{\textbf{w}}^{*}\nu^{*}_{1}=0.\]

\noindent From the above result, we see that $\nu^{*} \in \tilde{\mathcal{K}}$. On the other hand, we see that $F_{\nu^{*}}: \left[L,U\right] \rightarrow \left[0, 1 \right]$ and $F_{\nu^{*}}(x)$ is $C^{2}((L,U))$. Remember that we assumed $c^{net}+r+\delta(\nu) <0$ $\forall \nu \in \tilde{\mathcal{K}}$. Therefore, when the value of $\nu^{*}$ is given by (\ref{3pro19}), we have $-r > c^{net} > -\frac{D}{K}-r$ and, consequently, $\alpha < 0$. This, in turn, gives $\frac{\partial}{\partial x}F_{\nu^{*}}>0$, $\frac{\partial^{2}}{\partial x^{2}}F_{\nu^{*}}<0$. 

\noindent Next, we substitute the maximizer into the wealth process. As a result, we obtain, after rearranging the terms, for $t < \tau_{L}^{w^{*}} \wedge \tau_{U}^{w^{*}}$
\begin{equation}
X_{\nu^{*}}^{*}(t)=\begin{cases} X(0)\exp\left\lbrace \left[(r+c^{net})\left(-\frac{1}{\alpha}-1\right)\right]t-\frac{1}{\alpha}\zeta^{'} B(t)\right\rbrace & \mbox{if}\,\, -\frac{1}{\alpha}1^{'}_{N}\textit{\textbf{w}}_{o} < 1; \\ X(0)\exp\left\lbrace \left[c^{net}+r+\frac{D}{K}-\frac{1}{2K}\right]t-\frac{1}{\alpha}\zeta^{'}_{\nu^{*}}B(t) \right\rbrace & \mbox{if}\,\, -\frac{1}{\alpha}1^{'}_{N}\textit{\textbf{w}}_{o} \geq 1. \label{3pro25} \end{cases} 
\end{equation}

\noindent Notice that the value of $\alpha$ is different based on the value of the investment strategy. As we showed previously, we have  
\begin{equation}
\alpha = \begin{cases} \frac{\|\zeta\|^{2}}{2(r+c^{net})} & \mbox{if}\,\, -\frac{1}{\alpha}1^{'}_{N}\textit{\textbf{w}}_{o} < 1; \\ -2(K(r+c^{net})+D) & \mbox{if}\,\, -\frac{1}{\alpha}1^{'}_{N}\textit{\textbf{w}}_{o} \geq 1. \label{3pro26} \end{cases} 
\end{equation}

\noindent By using (\ref{3pro25}) we write
\begin{equation}
\mathbb{P}\left(X_{\nu^{*}}^{*}(t) \geq U \right) =\begin{cases} \Phi \left(\frac{\ln\left(\frac{x}{U}\right)+ \left[(r+c^{net})\left(-\frac{1}{\alpha}-1\right)\right]t}{\sqrt{\frac{1}{\alpha^{2}}\|\zeta\|^{2} t}}\right)& \mbox{if}\,\, -\frac{1}{\alpha}1^{'}_{N}\textit{\textbf{w}}_{o} < 1; \\ \Phi \left(\frac{\ln\left(\frac{x}{U}\right)+ \left[c^{net}+r+\frac{D}{K}-\frac{1}{2K}\right]t}{\sqrt{\frac{1}{\alpha^{2}}\|\zeta_{\nu^{*}}\|^{2}t}}\right) & \mbox{if}\,\, -\frac{1}{\alpha}1^{'}_{N}\textit{\textbf{w}}_{o} \geq 1. \label{3pro27} \end{cases}
\end{equation}

\noindent When $-\frac{1}{\alpha}1^{'}_{N}\textit{\textbf{w}}_{o} < 1$, the drift of the stochastic differential equation corresponding to $X_{\nu^{*}}(\cdot)$ is positive if $\alpha <-1$. Then, $\mathbb{P}\left(X_{\nu^{*}}^{*}(t) \geq U \right) \rightarrow 1$ as $t \rightarrow \infty$, and when $-1 < \alpha < 0$, $\mathbb{P}\left(X_{\nu^{*}}^{*}(t) \geq U \right) \rightarrow 0$ as $t \rightarrow \infty$. The same arguments hold for the term in the second case as well. That is, since we are considering the case $-\frac{1}{\alpha}1^{'}_{N}\textit{\textbf{w}}_{o} \geq 1$, the value of $\alpha$ is equal to $-2(K(r+c^{net})+D)$. If $\alpha < -1$, then $-2(K(r+c^{net})+D)$. This in turn implies that $c^{net}+r+\frac{D}{K}-\frac{1}{2K} < 0$.  Therefore, the drift of the term in the second case of (\ref{3pro27}) is negative. On the other hand, when $-1 < \alpha < 0$ the drift of . As a result, we have $\tau_{U}^{w^{*}} < \infty$ or $\tau_{L}^{w^{*}} < \infty$ almost surely. Then, we can write
\begin{equation}
\int_{0}^{\tau_{L}^{w^{*}} \wedge \tau_{U}^{w^{*}}}\|\textit{\textbf{w}}^{*}(X^{*}_{\nu^{*}}(s))\|^{2}ds=\begin{cases} \frac{1}{\alpha^{2}}\|\textit{\textbf{w}}_{o}\|^{2} (\tau_{L}^{w^{*}} \wedge \tau_{U}^{w^{*}}) < \infty & \mbox{if}\,\, -\frac{1}{\alpha}1^{'}_{N}\textit{\textbf{w}}_{o} < 1; \\ \frac{1}{K}\Sigma^{-1}(\tau_{L}^{w^{*}} \wedge \tau_{U}^{w^{*}}) < \infty & \mbox{if}\,\, -\frac{1}{\alpha}1^{'}_{N}\textit{\textbf{w}}_{o} \geq 1 \label{3pro28} \end{cases}
\end{equation}

\noindent almost surely, giving us $\textit{\textbf{w}}^{*} \in \mathcal{A}_{c}(x)$. Next, we fix a point $x \in (0,L)$ and introduce
\[\tau^{*}=\tau^{w}_{L} \wedge \tau^{w}_{U}\]

\noindent along with
\begin{equation}
\tau_{n}^{w}=\tau^{*} \wedge \inf \{t > 0 \mid \int_{0}^{t}\|\textit{\textbf{w}}(X^{w}_{\nu^{*}}(s))\|^{2}ds = n\} \quad \mbox{for} \quad n \in \mathbb{N}. \nonumber
\end{equation}

\noindent Then, we choose an arbitrary strategy $\textit{\textbf{w}} \in \mathcal{A}_{c}(x)$, and from It\^{o}'s formula we write by inserting $X^{w}_{\nu^{*}}(\cdot)$ into the function $F_{\nu^{*}}$
\begin{equation}
F_{\nu^{*}}\left(X^{w}_{\nu^{*}}(\tau_{n}^{w})\right)=F_{\nu^{*}}\left(x\right)+\int_{0}^{\tau_{n}^{w}}\{\mathcal{L}^{w}F_{\nu^{*}}\left(X^{w}_{\nu^{*}}(s)\right)\}ds +\int_{0}^{\tau_{n}^{w}}X^{w}_{\nu^{*}}(s)\textit{\textbf{w}}^{'}(X^{w}_{\nu^{*}}(s))\sigma\frac{\partial F_{\nu^{*}}}{\partial x}(X^{w}_{\nu^{*}}(s))dB(s). \label{3pro29}
\end{equation}

\noindent When we take the expectation of the above term, the stochastic integral vanishes. We can see this by substituting the value function in (\ref{3pro16}) to the following:
\begin{eqnarray}
&&\mathbb{E}_{x}\left[\int_{0}^{\tau_{n}^{w}}\left\|X^{w}_{\nu^{*}}(s)\textit{\textbf{w}}^{'}(X^{w}_{\nu^{*}}(s))\sigma\frac{\partial F_{\nu^{*}}}{\partial x}(X^{w}_{\nu^{*}}(s))\right\|^{2}ds \right] \nonumber \\
&\leq &\Sigma (1+\alpha)^{2}\frac{U^{2\alpha+2}}{(L^{1+\alpha}-U^{1+\alpha})^{2}}\mathbb{E}_{x}\left[\int_{0}^{\tau_{n}^{w}}\|\textit{\textbf{w}}^{'}(X^{w}_{\nu^{*}}(s))\|^{2}ds\right] \nonumber \\
& \leq & \Sigma (1+\alpha)^{2}\frac{U^{2\alpha+2}}{(L^{1+\alpha}-U^{1+\alpha})^{2}} n < \infty . \label{3veq20} 
\end{eqnarray}

\noindent On the other hand, notice that the function $F_{\nu^{*}}(x)$ solves the HJB equation. Thus, we have, $\forall \textit{\textbf{w}} \in \mathbb{R}^{N}$ and for each $s$ and $\mathbb{P}$-a.s., the inequality 
\begin{equation}
\{\mathcal{L}^{w}F_{\nu^{*}}\left(X^{w}_{\nu^{*}}(s)\right)\} \leq 0.\label{3veq18}
\end{equation}

\noindent From (\ref{3veq20}) and (\ref{3veq18}), we obtain the inequality
\begin{equation}
F_{\nu^{*}}\left(x\right) \geq \mathbb{E}_{x}\left[F_{\nu^{*}}\left(X^{w}_{\nu^{*}}(\tau_{n}^{w})\right)\right].\label{3veq21}
\end{equation}

\noindent Notice that $F_{\nu^{*}}(x)$ is a probability function. That is, $F_{\nu^{*}}(x) \in [0,1]$ for $x \in [L,U]$. In addition, we have as $n \rightarrow \infty$, $\tau^{w}_{n} \rightarrow \tau^{*}$. Therefore, from the dominated convergence theorem, we have as $n \rightarrow \infty$
\[\mathbb{E}_{x}\left[F_{\nu^{*}}\left(X^{w}_{\nu^{*}}(\tau_{n}^{w})\right)\right] \rightarrow \mathbb{E}_{x}\left[F_{\nu^{*}}\left(X^{w}_{\nu^{*}}(\tau^{*})\right)\right].\]

\noindent We can then write the inequality
\begin{eqnarray}
F_{\nu^{*}}\left(x\right) &\geq & \mathbb{E}_{x}\left[F_{\nu^{*}}\left(X^{w}_{\nu^{*}}(\tau^{*})\right)\right] \nonumber \\
&=&\mathbb{P}_{x}\left(\tau^{w}_{U} < \tau^{w}_{L}\right), \label{3veq22}
\end{eqnarray}

\noindent where we used the boundary conditions for the last line. When we take the supremum over admissible strategies, we have
\[F_{\nu^{*}}\left(x\right)\geq \sup_{\textit{\textbf{w}} \in \mathcal{A}_{c}}\mathbb{P}_{x}\left(\tau^{w}_{U} < \tau^{w}_{L}\right)=F_{\nu^{*}}\left(x\right),\]

\noindent showing us that $\textit{\textbf{w}}^{*}(x)$ is the optimal portfolio strategy, $X^{*}_{\nu^{*}}(t)$ is the optimal wealth process, and $F_{\nu^{*}}(x)$ is the optimal value function. In fact, with $\textit{\textbf{w}}^{*}(X^{*}_{\nu^{*}}(s))$ the supremum of the term $\{\mathcal{L}^{w}F\left(X^{w}_{\nu^{*}}(s)\right)\}$ is attained. In addition, we can show as we did in (\ref{3veq20}) that the stochastic integral vanishes. Then, the inequality in (\ref{3veq21}) becomes an equality and the results follow similarly proving the optimality of the results. 
\end{proof}

\section{Maximising/Minimising The Expected Time.} \label{3problem2} In addition to the strategy that maximises the probability of hitting a target, the investor may follow a strategy that maximises the time to hit the lower barrier $L$ when $U=\infty$, or minimises the time to reach the upper barrier $U$ when $L=0$. 

We know that the log-optimal growth strategy is the strategy that maximises the drift value of the investor's portfolio process. Then, by investing more or less than the log-optimal amount, an investor can have the chance to increase the probability of hitting an upper barrier $U$ without first hitting the lower barrier $L$. However, we wonder whether an investor can also maximise the time to survive or minimise the time to achieve a goal by investing in amounts different than the log-optimal strategy. To this end, we define the objective to maximise the time to hit the lower boundary $L$ with
\[\textit{\underbar{F}}_{\nu}(x)=\sup_{\textit{\textbf{w}} \in \mathcal{A}_{\nu}} \mathbb{E}_{x}\left[\tau_{L}^{w}\right]\]

\noindent and the objective to minimise the time to reach the upper boundary $U$ with
\[\bar{F}_{\nu}(x)=\inf_{\textit{\textbf{w}} \in \mathcal{A}_{\nu}} \mathbb{E}_{x}\left[\tau_{U}^{w}\right]\]

\begin{proposition} \label{prop2}
Let $\beta_{\nu} \in \mathbb{R} \setminus \{0\}$ be given by (\ref{3remark2}) and the portfolio process $\{X^{w}_{\nu}(t), 0 \leq t < \infty\}$ be the solution to the stochastic differential equation given in (\ref{3eq7}). Let also the optimal fictitious parameter be given by $\nu^{*}=\nu^{*}_{1}1_{N}$, and $\nu_{1}^{*}$ minimises
\[\frac{1}{2}\|\zeta+\nu_{1}\sigma^{-1}1_{N}\|^{2}-\nu_{1},\]

\noindent and is equal to\footnote{We use $D$ when specifying the cases. Notice that by the definition of $D$ we have $\zeta^{'}\sigma^{-1}1_{N}=\textit{\textbf{w}}_{o}^{'}1_{N}$. Therefore, $D < 1$ is equivalent to $\textit{\textbf{w}}_{o}^{'}1_{N}$.}
\begin{equation}
\nu^{*}=\begin{cases} 0_{N} & \mbox{if}\,\, D < 1; \\ \frac{1-D}{K}1_{N} & \mbox{if}\,\, D \geq 1 \label{3ieq26} \end{cases}
\end{equation}

\noindent with $D=\zeta^{'}\sigma^{-1}1_{N}$, $K=1_{N}^{'}\Sigma^{-1}1_{N}$. Then, $\beta_{\nu^{*}}$ is
\begin{equation}
\beta_{\nu^{*}}=\begin{cases} r+c^{net}+\frac{\|\zeta\|^{2}}{2} & \mbox{if}\,\, D < 1; \\ r+c^{net}-\frac{1}{2K}+\frac{D}{K} & \mbox{if}\,\, D \geq 1. \label{3ieq26b} \end{cases}
\end{equation}

\noindent (i) Suppose that $\beta_{\nu^{*}} < 0$. Then, the optimal value function for this problem is given by
\begin{equation}
\underbar{F}_{\nu^{*}}\left(x\right)=\frac{1}{\vert \beta_{\nu^{*}} \vert}\ln\left(\frac{x}{L}\right) \quad \mbox{for} \,\, x \in [L,\infty). \label{3ieq25}
\end{equation}

\noindent (ii) Next, suppose that $\beta_{\nu^{*}} > 0$. Then, the optimal value function for this problem is given by
\begin{equation}
\bar{F}_{\nu^{*}}\left(x\right)=\frac{1}{\beta_{\nu^{*}}}\ln\left(\frac{U}{x}\right) \quad \mbox{for} \,\, x \in \left(0,U\right]. \label{3ieqi25}
\end{equation}

\noindent The optimal investment strategy in both cases when $L < x < U$ is
\begin{equation}
\textbf{w}^{*}(x)=\begin{cases} \textbf{w}_{o} & \mbox{if}\,\, D < 1; \\ (\sigma^{'})^{-1}\left(\zeta+\frac{1-D}{K}\sigma^{-1}1_{N}\right) & \mbox{if}\,\, D \geq 1, \label{3eq37} \end{cases}
\end{equation}  

\noindent and the optimal wealth process satisfies for $t < \tau_{L}^{w^{*}} \wedge \tau_{U}^{w^{*}}$
\begin{equation}
X^{*}_{\nu^{*}}(t)=\begin{cases} X(0)\exp\left\lbrace \beta_{\nu^{*}} t+\zeta^{'} B(t)\right\rbrace & \mbox{if}\,\, D < 1; \\ X(0)\exp\left\lbrace \beta_{\nu^{*}}t+\zeta_{\nu^{*}}^{'}B(t) \right\rbrace & \mbox{if}\,\, D \geq 1, \label{3eq37b} \end{cases}
\end{equation}

\noindent with $\zeta_{\nu^{*}}=\zeta + \frac{1-D}{K}\sigma^{-1}1_{N}$.
\end{proposition}

We see that the log-optimal strategy is optimal for maximising the survival time and for minimising the time to reach a goal. As explained in \citet{Browne3} this is due to the fact that the optimal growth strategy is the strategy that maximises the mean rate of return of the investment portfolio. This, in turn, maximises the compounding rate and lead to the minimization of reaching a goal and to the maximization of survival. However, the existence of the results depend on the sign of the favourability parameter $\beta_{\nu^{*}}$. Notice that when $\beta_{\nu^{*}}<0$ (unfavourable), log-optimal strategy is the investment strategy that maximises survival. On the other hand, when $\beta_{\nu^{*}}>0$ (favourable), log-optimal strategy is the investment strategy that minimises the time to reach a goal.

\begin{remark}
Given the specification of $\alpha$ in (\ref{thr3}), we are in the favourable market when $\alpha < -1$, and in the unfavourable market when $-1 < \alpha < 0$. Therefore, under the unfavourable condition the investor takes more risk to maximise the probability of hitting a target before default compared to the risk that she takes to maximise the expected time to hit the lower boundary. The situation is reversed when the conditions are favourable. That is, the investor takes less risk to minimise the probability of ruin than the risk she takes to achieve a goal. These claims are equivalent to those given in \citet{Browne3} for benchmarked strategies; in unfavourable games {\em bold play} maximises the probability of success and {\em timid play} maximises the expected time to play. However, in favourable games {\em timid play} minimises the probability of ruin and {\em bold play} minimises the time to achieve a goal.
\end{remark}

\begin{proof} We let $\textit{\underbar{F}}_{\nu}(x)$ be the function given by (\ref{3ieq25}) with $\nu$ instead of $\nu^{*}$. We see that $\textit{\underbar{F}}_{\nu}(x)$ is continuous on $\left[L,\infty\right)$ and a $C^{2}\left(\left(L,\infty\right)\right)$ function with $\frac{\partial}{\partial x}\textit{\underbar{F}}_{\nu} > 0$, $\frac{\partial^{2}}{\partial x^{2}}\textit{\underbar{F}}_{\nu}<0$. In addition, it satisfies the boundary condition $\textit{\underbar{F}}_{\nu}(L)=0$. Then, the HJB equation over all strategies is for $\nu \in \tilde{\mathcal{K}}$
\begin{equation}
1+\left(r+c^{net}+\delta(\nu)\right)x\frac{\partial}{\partial x}\textit{\underbar{F}}_{\nu}+\sup_{\textit{\textbf{w}}}\left\lbrace \textit{\textbf{w}}^{'}(\mu +\nu- r1_{N})x\frac{\partial}{\partial x}\textit{\underbar{F}}_{\nu}+\frac{1}{2}\textit{\textbf{w}}^{'}\Sigma\textit{\textbf{w}}x^{2}{\frac{\partial^{2}}{\partial x^{2}}\textit{\underbar{F}}_{\nu}}\right\rbrace =0. \label{3ip21}
\end{equation}

\noindent By using (\ref{3ip21}), we give the maximising $\textit{\textbf{w}}^{*}$ by
\begin{equation}
\textit{\textbf{w}}^{*}(x)=-(\sigma^{'})^{-1}\zeta_{\nu}\frac{\frac{\partial}{\partial x}\textit{\underbar{F}}_{\nu}}{x\frac{\partial^{2}}{\partial x^{2}}\textit{\underbar{F}}_{\nu}}.\label{3ip22}
\end{equation}

\noindent After substituting (\ref{3ip21}) into (\ref{3ip22}), we get the non-linear partial differential equation
\begin{equation}
1+\left(r+c^{net}+\delta(\nu)\right)x\frac{\partial}{\partial x}\textit{\underbar{F}}_{\nu}-\frac{1}{2}\|\zeta+\sigma^{-1}\nu\|^{2}\frac{\left(\frac{\partial}{\partial x}\textit{\underbar{F}}_{\nu}\right)^{2}}{\frac{\partial^{2}}{\partial x^{2}}\textit{\underbar{F}}_{\nu}} =0. \label{3ip23}
\end{equation}

\noindent When we substitute the value function $\textit{\underbar{F}}_{\nu}(x)$ into (\ref{3ip23}), we observe that the equality is satisfied. Then, the maximizer is given by
\begin{equation}
\textit{\textbf{w}}^{*}(x)=\begin{cases} \textit{\textbf{w}}_{o} & \mbox{if}\,\, D < 1; \\ (\sigma^{'})^{-1}\zeta_{\nu} & \mbox{if}\,\, D \geq 1. \label{3ip23b} \end{cases}
\end{equation}  

\noindent Next, we write by the application of It\^{o}'s formula
\begin{equation}
\textit{\underbar{F}}_{\nu}(X^{w}_{\nu}(t))=\textit{\underbar{F}}_{\nu}(X(0))-t-\frac{1}{\beta_{\nu}}\textit{\textbf{w}}^{'}_{o}\sigma B(t)  \quad \mbox{when} \,\, t \leq \tau_{L}^{w}.\label{nu42}
\end{equation}

\noindent Notice that $\beta_{\nu}$ is an increasing function of $\nu$. Therefore, $\nu^{*}$ is the minimizer for (\ref{nu42}) because $\textit{\underbar{F}}_{\nu^{*}}(x)\leq \textit{\underbar{F}}_{\nu}(x)$ $\forall \nu \in \tilde{\mathcal{K}}$. As a result, we find the optimal fictitious parameter by minimising $\beta_{\nu}$. In other words, we aim to find $\nu^{*}=\nu^{*}_{1}1_{N}$ that minimises the term
\[\frac{1}{2}\|\zeta+\nu_{1}\sigma^{-1}1_{N}\|^{2}-\nu_{1}.\]

\noindent So $\nu_{1}^{*}$ is given by
\begin{equation}
\nu_{1}^{*}=\begin{cases} 0 & \mbox{if}\,\, D < 1; \\ \frac{1-D}{K} & \mbox{if}\,\, D \geq 1. \label{3ip24b} \end{cases}
\end{equation}

\noindent When we substitute $(1-D)/K$ into $\|\zeta+\nu_{1}\sigma^{-1}1_{N}\|^{2}$, we have
\[\frac{1}{2}\|\zeta+\frac{1-D}{K}\sigma^{-1}1_{N}\|^{2}=\frac{1}{2K}. \]

\noindent Therefore, the favourability parameter $\beta_{\nu^{*}}$ under the constrained market scenario is given by (\ref{3ieq26b}), and the maximizer is specified as
\begin{equation}
\textit{\textbf{w}}^{*}(x)=\begin{cases} \textit{\textbf{w}}_{o} & \mbox{if}\,\, D < 1; \\ (\sigma^{'})^{-1}\left(\zeta+\frac{1-D}{K}\sigma^{-1}1_{N}\right) & \mbox{if}\,\, D \geq 1. \label{maximiser2} \end{cases}
\end{equation} 

\noindent Next, we verify that (\ref{3ieq25}) is the optimal value function, and (\ref{maximiser2}) is the optimal investment strategy. To do this, we first check if $\textit{\textbf{w}}^{*} \in \mathcal{K}$. We have, when $D \geq 1$
\begin{eqnarray}
1^{'}_{N}\textit{\textbf{w}}^{*}(x)&=&1^{'}_{N}(\sigma^{'})^{-1}\left(\zeta+\frac{1-D}{K}\sigma^{-1}1_{N}\right) \nonumber \\
&=&D+1-D=1. \label{checkK}
\end{eqnarray}

\noindent The maximising strategy is $\textit{\textbf{w}}_{o}$ when $D < 1$. Thus, $\textit{\textbf{w}}^{*} \in \mathcal{K}$. Then, we can see that $-\nu^{*}_{1}+1_{N}^{'}\textit{\textbf{w}}^{*}\nu^{*}_{1}=0$, giving us $\nu^{*} \in \tilde{\mathcal{K}}$. On the other hand, when the maximizer is substituted into the wealth process, we obtain 
\begin{equation}
X^{*}_{\nu^{*}}(t)=\begin{cases} X(0)\exp\left\lbrace \beta_{\nu^{*}} t+\zeta^{'} B(t)\right\rbrace & \mbox{if}\,\, D < 1; \\ X(0)\exp\left\lbrace \beta_{\nu^{*}}t+\zeta_{\nu^{*}}^{'}B(t) \right\rbrace & \mbox{if}\,\, D \geq 1, \label{maximiser3} \end{cases}
\end{equation}

\noindent where $\beta_{\nu^{*}}$ is given by
\begin{equation}
\beta_{\nu^{*}}=\begin{cases} r+c^{net}+\frac{\|\zeta\|^{2}}{2} & \mbox{if}\,\, D < 1; \\ r+c^{net}-\frac{1}{2K}+\frac{D}{K} & \mbox{if}\,\, D \geq 1. \label{3iiiieq26b} \end{cases}
\end{equation}

\noindent We are in unfavourable markets. Therefore, $\beta_{\nu^{*}} < 0$ and $\mathbb{P}(X^{*}_{\nu^{*}}(t) \geq L) \rightarrow 0$ as $t \rightarrow \infty$. This gives $\tau^{w^{*}}_{L} < \infty$ almost surely. It follows that
\begin{equation}
\int_{0}^{\tau_{L}^{w^{*}}}\|\textit{\textbf{w}}^{*}(X_{\nu^{*}}^{*}(s))\|^{2}ds=\begin{cases} \|\textit{\textbf{w}}_{o}\|^{2} \tau_{L}^{w^{*}} < \infty & \mbox{if}\,\, D < 1; \\ \frac{1}{K}\Sigma^{-1}\tau_{L}^{w^{*}} < \infty & \mbox{if}\,\, D \geq 1 \label{maximiser4} \end{cases}
\end{equation}

\noindent almost surely. As a result, $\textit{\textbf{w}}^{*} \in \mathcal{A}_{c}(x)$. Next, we fix $x \in (L,\infty)$ and introduce the function
\begin{equation}
M_{\nu^{*}}\left(t,X^{w}_{\nu^{*}}(t)\right)=\textit{\underbar{F}}_{\nu^{*}}\left(X^{w}_{\nu^{*}}(t)\right)+t \label{33veq15}
\end{equation}

\noindent along with the stopping time
\[\tau_{n}^{w}=\tau_{L}^{w} \wedge \inf\{t > 0 \mid \int_{0}^{t} \|\textit{\textbf{w}}(X^{w}_{\nu^{*}}(s))\|^{2}ds = n\} \quad \mbox{for} \quad n \in \mathbb{N}\]

\noindent Then, we choose an arbitrary strategy $\textit{\textbf{w}} \in \mathcal{A}_{c}(x)$, and from It\^{o}'s formula we write by inserting $X^{w}_{\nu^{*}}(\cdot)$ into the function $F_{\nu^{*}}$
\begin{eqnarray}
M_{\nu^{*}}\left(\tau_{n}^{w},X^{w}_{\nu^{*}}(\tau_{n}^{w})\right)&=&\textit{\underbar{F}}_{\nu^{*}}\left(x\right)+\tau_{n}^{w}+\int_{0}^{\tau_{n}^{w}}\{\mathcal{L}^{w}\textit{\underbar{F}}_{\nu^{*}}\left(X^{w}_{\nu^{*}}(s)\right)\}ds \nonumber \\
& & \, +\,\int_{0}^{\tau_{n}^{w}}X^{w}_{\nu^{*}}(s)\textit{\textbf{w}}^{'}(X^{w}_{\nu^{*}}(s))\sigma \frac{\partial \textit{\underbar{F}}_{\nu^{*}}}{\partial x}(X^{w}_{\nu^{*}}(s))dB(s). \nonumber \\ \label{inewproof2}
\end{eqnarray} 

\noindent When the expectation of the above terms is taken, the stochastic integral vanishes. This can be observed by using the value function $\textit{\underbar{F}}_{\nu^{*}}(X_{\nu^{*}}^{w}(t))$ in the following:
\begin{eqnarray}
&&\mathbb{E}_{x}\left[\int_{0}^{\tau_{n}^{w}}\left\|X^{w}_{\nu^{*}}(s)\textit{\textbf{w}}^{'}(X^{w}_{\nu^{*}}(s))\sigma \frac{\partial \textit{\underbar{F}}_{\nu^{*}}}{\partial x}(X^{w}_{\nu^{*}}(s))\right\|^{2} ds \right] \nonumber \\
&\leq & \Sigma\frac{1}{\beta_{\nu^{*}}^{2}}\mathbb{E}_{x}\left[\int_{0}^{\tau_{n}^{w}}\left\|\textit{\textbf{w}}^{'}(X^{w}_{\nu^{*}}(s))\right\|^{2}ds \right] \nonumber \\
& \leq &\Sigma\frac{1}{\beta_{\nu^{*}}^{2}}n  < \infty . \label{inewproof3}
\end{eqnarray}

\noindent We know that $\textit{\underbar{F}}_{\nu^{*}}\left(x\right)$ solves the HJB equation. Then, $\forall \textit{\textbf{w}} \in \mathbb{R}^{N}$, we have for each $s$ and $\mathbb{P}$-a.s.
\begin{equation}
1+\mathcal{L}^{w}\textit{\underbar{F}}_{\nu^{*}}\left(X^{w}_{\nu^{*}}(s)\right) \leq 0. \label{inewproof4}
\end{equation}

\noindent By using (\ref{inewproof3}) and (\ref{inewproof4}), we obtain the inequality
\begin{equation}
\textit{\underbar{F}}_{\nu^{*}}\left(x\right) \geq \mathbb{E}_{x}\left[M_{\nu^{*}}\left(\tau_{n}^{w},X^{w}_{\nu^{*}}(\tau_{n}^{w})\right)\right]. \label{inewproof5}
\end{equation}

\noindent We have $\textit{\underbar{F}}_{\nu^{*}}\left(x\right) \leq \frac{1}{\vert \beta_{\nu^{*}} \vert}\ln(x/L)+ \mathcal{C} < \infty$ for some constant $\mathcal{C}\geq 0$. In addition, observe that as $n \rightarrow \infty$, $\tau_{n}^{w} \rightarrow \tau_{L}^{w}$. Therefore, from the dominated convergence theorem, we have as $n \rightarrow \infty$
\begin{equation*}
\mathbb{E}_{x}\left[M_{\nu^{*}}\left(\tau_{n}^{w},X^{w}_{\nu^{*}}(\tau_{n}^{w})\right) \right] \rightarrow \mathbb{E}_{x}\left[M_{\nu^{*}}\left(\tau_{L}^{w},X^{w}_{\nu^{*}}(\tau_{L}^{w})\right)\right]. 
\end{equation*}

\noindent Thus, the inequality in (\ref{inewproof5}) becomes
\begin{eqnarray}
\textit{\underbar{F}}_{\nu^{*}}\left(x\right) &\geq & \mathbb{E}_{x}\left[M_{\nu^{*}}\left(\tau_{L}^{w},X^{w}_{\nu^{*}}(\tau_{L}^{w})\right)\right] \nonumber \\
&=&\mathbb{E}_{x}\left[\tau_{L}^{w}\right]. \label{inewproof5b}
\end{eqnarray}

\noindent When the supremum is taken over all admissible strategies we obtain
\[\textit{\underbar{F}}_{\nu^{*}}\left(x\right) \geq \sup_{\textit{\textbf{w}} \in \mathcal{A}_{c}} \mathbb{E}_{x}\left[\tau_{L}^{w} \right]=\textit{\underbar{F}}_{\nu^{*}}\left(x\right).\]

\noindent In fact, when the maximising strategy $\textit{\textbf{w}}^{*}(X^{*}_{\nu^{*}}(s))$ is chosen, we have for each $s$ and $\mathbb{P}$-a.s.
\begin{equation}
1+\mathcal{L}^{w^{*}}\textit{\underbar{F}}_{\nu^{*}}\left(X^{*}_{\nu^{*}}(s)\right) = 0 \label{inewproof7}
\end{equation}

\noindent In addition, we can show similarly as we did in (\ref{inewproof3}) that the stochastic integral vanishes with $\textit{\textbf{w}}^{*}(X^{*}_{\nu^{*}}(s))$.  Thus, the inequality in (\ref{inewproof5}) becomes an equality and we obtain
\begin{equation}
\textit{\underbar{F}}_{\nu^{*}}\left(x\right) = \mathbb{E}_{x}\left[\tau_{L}^{w^{*}}\right] \label{inewproof8}
\end{equation}

\noindent showing that $\textit{\textbf{w}}^{*}(x)$ is the optimal strategy, $X^{*}_{\nu^{*}}(t)$ is the optimal wealth process, and $\textit{\underbar{F}}_{\nu^{*}}(x)$ is the optimal value function.
\\

\noindent Notice that the minimization problem can be written as $\bar{F}(x)=-\sup_{\textit{\textbf{w}} \in \mathcal{A}}\mathbb{E}\left[-\tau_{U}^{w}\mid X(0)=x\right]$. Then, we can apply the approach of the maximization problem to the function $\hat{F}(x)=-\bar{F}(x)$ for the proof of the results.
\end{proof}

\section{Maximising and Minimising The Expected Discounted Reward.} \label{3problem3} In the previous sections we aimed to find the optimal investment strategies that maximise the probability of hitting an upper barrier before default and strategies that either maximise the time to hit the lower barrier or minimise the time to reach the upper barrier. This section is related to finding the optimal investment strategy of an investor expecting a cash inflow or outflow in the future when her wealth reaches a certain level. For example, when the wealth of the investor reaches an upper level $U$ she may be paid a fixed amount. In this case, the investor will seek to invest optimally in order to maximise the discounted value of the amount that she expects to receive when $U$ is reached. On the other hand, the minimization problem is related to the penalty that the investor is supposed to pay when her wealth level will reach a lower boundary $L$. In this case, the investor would seek to find a strategy that minimises the expected discounted value of the penalty she is expected to pay when the wealth level hits the lower barrier $L$. 

Given the aims of the investor above, we define the objective to maximise the expected discounted reward from reaching the upper barrier by 
\[\bar{G}_{\nu}(x)=\sup_{\textit{\textbf{w}} \in \mathcal{A}_{\nu}} \mathbb{E}_{x}\left[e^{-\rho\tau_{U}^{w}}\right],\]

\noindent where $\rho > 0$ is the constant discount rate. On the other hand, the objective to minimise the expected discounted reward from hitting the lower barrier is defined by
\[\textit{\underbar{G}}_{\nu}(x)=\inf_{\textit{\textbf{w}} \in \mathcal{A}_{\nu}} \mathbb{E}_{x}\left[e^{-\rho\tau_{L}^{w}}\right].\]

We then present our solutions in the next proposition.

\begin{proposition} \label{prop3}
Let the portfolio process $\{X^{w}_{\nu}(t), 0 \leq t < \infty\}$ be the solution to the stochastic differential equation given in (\ref{3eq7}). Let also $\nu_{1}^{*}$ be the minimizer of the term
\[\frac{d}{2(d+1)}\|\zeta+\nu_{1}\sigma^{-1}1_{N}\|^{2}-d\nu_{1}.\]

\noindent (i) For the maximization problem, the optimal fictitious parameter is\footnote{As in the first problem, we state the cases by using the value of the optimal investment strategy given by the first line in (\ref{3ieq42a}). That is, $d^{-}$ in $\frac{1}{1+d^{-}}1_{N}^{'}\textit{\textbf{w}}_{o} \geq 1$ is given by the first line of (\ref{3ieq44a}). The same argument holds for the minimization problem as well.}
\begin{equation}
\nu^{*}=\begin{cases} 0_{N} & \mbox{if}\,\, \frac{1}{1+d^{-}}1_{N}^{'}\textit{\textbf{w}}_{o} < 1; \\ \frac{d^{-}+1-D}{K}1_{N} & \mbox{if}\,\, \frac{1}{1+d^{-}}1_{N}^{'}\textit{\textbf{w}}_{o}\geq 1 \label{3ieq26} \end{cases}
\end{equation}

\noindent with $D=\zeta^{'}\sigma^{-1}1_{N}$, $K=1_{N}^{'}\Sigma^{-1}1_{N}$, and $-1 < d^{-} < 0$ is given by, for $k=\|\zeta\|^{2}/2$,
\begin{equation}
d^{-}= \begin{cases} \frac{1}{2(r+c^{net})}\left[-\left(r+c^{net}+k+\rho\right)+\sqrt{\left(r+c^{net}+k+\rho\right)^{2}-4(r+c^{net})\rho}\right] & \mbox{if}\,\frac{1}{1+d^{-}}1_{N}^{'}\textit{\textbf{w}}_{o} < 1; \\ \frac{1}{2}\left[-\left[1-2\left(K(r+c^{net})+D\right)\right]-\sqrt{\left[1-2\left(K(r+c^{net})+D\right)\right]^{2}+8K\rho}\right] & \mbox{if}\,\frac{1}{1+d^{-}}1_{N}^{'}\textit{\textbf{w}}_{o} \geq 1, \label{3ieq44a}\end{cases} 
\end{equation}

\noindent If $r+c^{net}+\frac{D}{K}-\frac{1}{2K}>0$ and the condition 
\begin{equation}
\left[\left(r+c^{net}\right)+\frac{\|\zeta\|^{2}}{1+d^{-}}\left(1-\frac{1}{2(1+d^{-})}\right)\right] > 0 \label{condition}
\end{equation}

\noindent holds then, we are in the favourable markets, and the optimal value function is equal to 
\begin{equation}
\bar{G}_{\nu^{*}}(x)=\left(\frac{x}{U}\right)^{-d^{-}} \quad \mbox{for} \,\, x \in \left[0,U\right]. \label{3ieq40b}
\end{equation}

\noindent Furthermore, the optimal investment strategy for $0 < x < U$ is
\begin{equation}
\textbf{w}^{*}(x)= \begin{cases} \frac{1}{1+d^{-}}\textit{\textbf{w}}_{o} & \mbox{if}\,\, \frac{1}{1+d^{-}}1_{N}^{'}\textit{\textbf{w}}_{o} < 1; \\ \frac{1}{1+d^{-}}(\sigma^{'})^{-1}\left(\zeta+\frac{d^{-}+1-D}{K}\sigma^{-1}1_{N}\right) & \mbox{if}\,\, \frac{1}{1+d^{-}}1_{N}^{'}\textit{\textbf{w}}_{o} \geq 1,\label{3ieq42a}\end{cases} 
\end{equation}

\noindent and the optimal wealth process for $t < \tau^{w^{*}}_{U}$ is
\begin{equation}
X^{*}_{\nu^{*}}(t)=\begin{cases} X(0)\exp\left\lbrace \left[\left(r+c^{net}\right)+\frac{\|\zeta\|^{2}}{1+d^{-}}\left(1-\frac{1}{2(1+d^{-})}\right)\right]t+\frac{1}{1+d^{-}}\zeta^{'}B(t)\right\rbrace &  \mbox{if}\,\, \frac{1}{1+d^{-}}1_{N}^{'}\textit{\textbf{w}}_{o} < 1; \\
X(0)\exp\left\lbrace \beta_{\nu^{*}}t+\frac{1}{1+d^{-}}\zeta_{\nu^{*}}^{'}B(t) \right\rbrace & \mbox{if}\,\, \frac{1}{1+d^{-}}1_{N}^{'}\textit{\textbf{w}}_{o} \geq 1
\label{444maximiserz3} \end{cases}
\end{equation}

\noindent with $\zeta_{\nu^{*}}=\zeta + \frac{d^{-}+1-D}{K}\sigma^{-1}1_{N}$.

\noindent (ii) For the minimization problem, we consider the unfavourable markets. Thus, $\beta_{\nu} < 0$ $\forall \nu \in \tilde{\mathcal{K}}$. In this case, the optimal fictitious parameter is
\begin{equation}
\nu^{*}=\begin{cases} 0_{N} & \mbox{if}\,\,\frac{1}{1+d^{-}}1_{N}^{'}\textbf{w}_{o} < 1; \\ \frac{d^{+}+1-D}{K}1_{N} & \mbox{if}\,\, \frac{1}{1+d^{+}}1_{N}^{'}\textbf{w}_{o} \geq 1, \label{3ieq26} \end{cases}
\end{equation}

\noindent where $d^{+} > 0$ is given by
\begin{equation}
d^{+}= \begin{cases} \frac{1}{2(r+c^{net})}\left[-\left(r+c^{net}+k+\rho\right)-\sqrt{\left(r+c^{net}+k+\rho\right)^{2}-4(r+c^{net})\rho}\right] & \mbox{if}\,\frac{1}{1+d^{+}}1_{N}^{'}\textbf{w}_{o} < 1; \\ \frac{1}{2}\left[-\left[1-2\left(K(r+c^{net})+D\right)\right]+\sqrt{\left[1-2\left(K(r+c^{net})+D\right)\right]^{2}+8K\rho}\right] & \mbox{if}\,\frac{1}{1+d^{+}}1_{N}^{'}\textbf{w}_{o} \geq 1, \label{3ieq44}\end{cases} 
\end{equation}

\noindent Then, the optimal value function is equal to 
\begin{equation}
\underbar{G}_{\nu^{*}}(x)=\left(\frac{x}{L}\right)^{-d^{+}} \quad \mbox{for} \,\, x \in \left[L,\infty\right). \label{3ieq46}
\end{equation}

\noindent Furthermore, the optimal investment strategy satisfies for $L < x < \infty$
\begin{equation}
\textbf{w}^{*}(x)= \begin{cases} \frac{1}{1+d^{+}}\textbf{w}_{o} & \mbox{if}\,\, \frac{1}{1+d^{+}}1_{N}^{'}\textbf{w}_{o} < 1; \\ \frac{1}{1+d^{+}}(\sigma^{'})^{-1}\left(\zeta+\frac{d^{+}+1-D}{K}\sigma^{-1}1_{N}\right) & \mbox{if}\,\, \frac{1}{1+d^{+}}1_{N}^{'}\textbf{w}_{o} \geq 1, \label{3ieq42}\end{cases} 
\end{equation}

\noindent and the optimal wealth process $t < \tau^{w^{*}}_{L}$ is
\begin{equation}
X^{*}_{\nu^{*}}(t)=\begin{cases} X(0)\exp\left\lbrace \left[\left(r+c^{net}\right)+\frac{\|\zeta\|^{2}}{1+d^{+}}\left(1-\frac{1}{2(1+d^{+})}\right)\right]t+\frac{1}{1+d^{+}}\zeta^{'} B(t)\right\rbrace & \mbox{if}\,\, \frac{1}{1+d^{+}}1_{N}^{'}\textbf{w}_{o} < 1; \\
X(0)\exp\left\lbrace \beta_{\nu^{*}}t+\frac{1}{1+d^{+}}\zeta_{\nu^{*}}^{'}B(t) \right\rbrace & \mbox{if}\,\, \frac{1}{1+d^{+}}1_{N}^{'}\textbf{w}_{o} \geq 1
\label{444maximiserz23} \end{cases}
\end{equation}

\noindent with $\zeta_{\nu^{*}}=\zeta+\frac{d^{+}+1-D}{K}1_{N}\sigma^{-1}1_{N}$.

\end{proposition}

Here as well, the optimal investment strategies are independent of the current wealth level and the barriers $U$ and $L$. As in the first problem, it is proportional to the log-optimal investment strategy. In addition, we observe that the reward maximization happens in the favourable markets and the penalty minimization happens in the unfavourable one. That is, under the favourable market an investor can find a reward maximising strategy. However, under the unfavourable market, the same investor should look for penalty minimising strategies instead. Therefore, when the markets are favourable there is no penalty to pay and when unfavourable there is no reward to gain. 

Furthermore, because $-1 < d^{-} <0$ and we are in the favourable market, we see that the investor takes more risk than she would with the log-optimal strategy when maximising the expected discounted reward. The situation is different under the minimization problem since $d^{+} > 0$, and we are in unfavourable market. In this case, the investor takes less risk in order to diminish the expected discounted value of the penalty. Therefore, in favourable games bold play {\em maximises the reward} and in unfavourable games timid play {\em minimises the penalty}.

\begin{proof} We start by assuming that $\bar{G}_{\nu}:\left[0,U\right] \rightarrow \left[0,1 \right]$ and $\bar{G}_{\nu}(x)$ is $C^{2}(\left(0,U)\right)$ with $\frac{\partial}{\partial x}\bar{G}_{\nu}>0$, $\frac{\partial^{2}}{\partial x^{2}}\bar{G}_{\nu}<0$. Then, the HJB equation over all strategies is for $\nu \in \tilde{\mathcal{K}}$
\begin{equation}
-\rho \bar{G}_{\nu}+\left(r+c^{net}+\delta(\nu)\right)x\frac{\partial}{\partial x}\bar{G}_{\nu}+\sup_{\textit{\textbf{w}}}\left\lbrace \textit{\textbf{w}}^{'}(\mu +\nu- r1_{N})x\frac{\partial}{\partial x}\bar{G}_{\nu}+\frac{1}{2}\textit{\textbf{w}}^{'}\Sigma\textit{\textbf{w}}x^{2}{\frac{\partial^{2}}{\partial x^{2}}\bar{G}_{\nu}}\right\rbrace =0 
\label{3ip31}
\end{equation}

\noindent subject to the boundary condition $\bar{G}_{\nu}(U)=1$. From (\ref{3ip31}), we find the maximizer $\textit{\textbf{w}}^{*}$ as
\begin{equation}
\textit{\textbf{w}}^{*}(x)=-(\sigma^{'})^{-1}\zeta_{\nu}\frac{\frac{\partial}{\partial x}\bar{G}_{\nu}}{x\frac{\partial^{2}}{\partial x^{2}}\bar{G}_{\nu}}.\label{3ip32}
\end{equation}

\noindent By substituting (\ref{3ip32}) into (\ref{3ip31}), we obtain the non-linear partial differential equation
\begin{equation}
-\rho \bar{G}_{\nu}+\left(r+c^{net}+\delta(\nu)\right)x\frac{\partial}{\partial x}\bar{G}_{\nu}-\frac{1}{2}\|\zeta+\sigma^{-1}\nu \|^{2}\frac{\left(\frac{\partial}{\partial x}\bar{G}_{\nu}\right)^{2}}{\frac{\partial^{2}}{\partial x^{2}}\bar{G}_{\nu}} =0. \label{3ip33}
\end{equation}

\noindent To solve the above equation we use a solution of the form $Ax^{-d}$ where $A$ is a constant. By substituting this function into (\ref{3ip33}) we obtain
\begin{equation}
Ax^{-d}\left(\rho+\left(r +c^{net}\right)d+\left[\frac{d}{2(d+1)}\|\zeta+\sigma^{-1}\nu\|^{2}+d\delta(\nu)\right]\right)=0. \label{3ip34}
\end{equation} 

\noindent The optimal fictitious parameter is $\nu^{*}=\nu^{*}_{1}1_{N}$, where $\nu^{*}_{1}$ is the minimizer of the term
\begin{equation}
\frac{d}{2(d+1)}\|\zeta+\nu_{1}\sigma^{-1}1_{N}\|^{2}-d\nu_{1}. \label{min1}
\end{equation}

\noindent Then, the values of $\nu_{1}^{*}$ are
\begin{equation}
\nu_{1}^{*}=\begin{cases} 0 & \mbox{if} \,\, 1^{'}_{N}\textit{\textbf{w}}^{*}(x) < 1;\\
\frac{d+1-D}{K} & \mbox{if} \,\,1^{'}_{N}\textit{\textbf{w}}^{*}(x) \geq 1. \label{1proof6} \end{cases}
\end{equation}

\noindent Furthermore,
\[\|\zeta+\nu_{1}^{*}\sigma^{-1}1_{N}\|^{2}=\frac{(d+1)^{2}}{K}.\]

\noindent By substituting the values in (\ref{1proof6}) and the above into (\ref{3ip34}), we find the quadratic equations as
\begin{eqnarray}
d^{2}\left(r+c^{net}\right)+d\left[r+c^{net}+k+\rho\right]+\rho &=& 0 \quad \mbox{if} \,\, 1^{'}_{N}\textit{\textbf{w}}^{*}(x) < 1; \nonumber\\
d^{2}+d\left[1-2\left(K(r+c^{net})+D\right)\right]-2K\rho &=& 0 \quad \mbox{if} \,\, 1^{'}_{N}\textit{\textbf{w}}^{*}(x) \geq 1. \label{1proof7}
\end{eqnarray}

\noindent The quadratic equations have two roots; $d^{+}$ and $d^{-}$. Therefore, the equation (\ref{3ip34}) admits two solutions $A_{1}x^{-d^{+}}$ and $A_{2}x^{-d^{-}}$, where $A_{1}, A_{2}$ are two constants. The roots are given by
\begin{equation}
d^{+}=\begin{cases} \frac{1}{2(r+c^{net})}\left[-\left(r+c^{net}+k+\rho\right)-\sqrt{\Delta}\right] & \mbox{if} \,\, 1^{'}_{N}\textit{\textbf{w}}^{*}(x) < 1;\\
\frac{1}{2}\left[-\left[1-2\left(K(r+c^{net})+D\right)\right]+\sqrt{\Delta_{b}}\right] & \mbox{if} \,\, 1^{'}_{N}\textit{\textbf{w}}^{*}(x) \geq 1, \label{1proof8} \end{cases}
\end{equation}

\noindent and

\begin{equation}
d^{-}=\begin{cases} \frac{1}{2(r+c^{net})}\left[-\left(r+c^{net}+k+\rho\right)+\sqrt{\Delta}\right] & \mbox{if} \,\, 1^{'}_{N}\textit{\textbf{w}}^{*}(x) < 1;\\
\frac{1}{2}\left[-\left[1-2\left(K(r+c^{net})+D\right)\right]-\sqrt{\Delta_{b}}\right] & \mbox{if} \,\, 1^{'}_{N}\textit{\textbf{w}}^{*}(x) \geq 1, \label{1proof9} \end{cases}
\end{equation}

\noindent where the discriminants are given by
\begin{eqnarray}
\Delta &=&\left(r+c^{net}+k+\rho\right)^{2}-4(r+c^{net})\rho; \label{3pr14a} \\
\Delta_{b}&=&\left[1-2\left(K(r+c^{net})+D\right)\right]^{2}+8K\rho . \label{3pr14b}
\end{eqnarray}

\noindent We can check that $d^{+}d^{-} < 0$. Therefore, the signs of the roots are different. We have $d^{-} < 0$. In order to use it for the maximization problem, we need to check that $d^{-} > -1$. We can see from the first term in (\ref{1proof9}) that this is the case. On the other hand, we assumed that $r+\delta(\nu)+c^{net} < 0$. By substituting the value of $\nu_{1}^{*}$ given in the second line of (\ref{1proof6}) into the aforementioned assumption and using $c^{net} > - (r+D/K)$ from the first problem\footnote{Remember that $\alpha \in (-\infty,0) \setminus {-1}$ and $\alpha=-2(K(r+c^{net})+D)$ when $-\frac{1}{\alpha}1^{'}_{N}\textbf{w}_{o} \geq 1$. It follows that $c^{net} > - \left(r+\frac{D}{K}\right)$.} we can confirm that the value of $d^{-}$ given by the second term in (\ref{1proof9}) is larger than $-1$ as well. Therefore, the value function for the maximization problem can be specified as 
\[\bar{G}_{\nu^{*}}(x)=\left(\frac{x}{U}\right)^{-d^{-}}\]

\noindent Notice that $\bar{G}_{\nu^{*}}(x) \leq \bar{G}_{\nu}(x)$ $\forall \nu \in \tilde{\mathcal{K}}$, confirming the minimization of the term (\ref{min1}) in order to find $\nu^{*}$.  Then, we write the maximising strategy as
\begin{equation}
\textit{\textbf{w}}^{*}(x)= \begin{cases} \frac{1}{1+d^{-}}\textit{\textbf{w}}_{o} & \mbox{if}\,\, \frac{1}{1+d^{-}}1_{N}^{'}\textit{\textbf{w}}_{o} < 1; \\ \frac{1}{1+d^{-}}(\sigma^{'})^{-1}\left(\zeta+\frac{d^{-}+1-D}{K}\sigma^{-1}1_{N}\right) & \mbox{if}\,\, \frac{1}{1+d^{-}}1_{N}^{'}\textit{\textbf{w}}_{o} \geq 1. \label{3ieqbbbb42}\end{cases} 
\end{equation}

\noindent Now, we show that the above results are optimal. First, we check if $\textit{\textbf{w}}^{*} \in \mathcal{K}$. We start by specifying when $\frac{1}{1+d^{-}}1_{N}^{'}\textit{\textbf{w}}_{o} \geq 1$
\begin{eqnarray}
1_{N}^{'}\textit{\textbf{w}}^{*}(x)&=&1_{N}^{'}\frac{1}{1+d^{-}}(\sigma^{'})^{-1}\left(\zeta+\frac{d^{-}+1-D}{K}\sigma^{-1}1_{N}\right) \nonumber \\
&=&\frac{1}{1+d^{-}}\left(D+d^{-}+1-D\right)=1. \label{maximiserz2}
\end{eqnarray}

\noindent On the other hand, when $\frac{1}{1+d^{-}}1_{N}^{'}\textit{\textbf{w}}_{o} < 1$, we have the first term in (\ref{3ieqbbbb42}) with $\nu^{*}=0$. Therefore, $\textit{\textbf{w}}^{*} \in \mathcal{K}$. For such $\textit{\textbf{w}}^{*}$ we can write
\[\delta(\nu^{*})+1_{N}^{'}\textit{\textbf{w}}^{*}\nu_{1}^{*}=0,\] 

\noindent making $\nu^{*} \in \tilde{\mathcal{K}}$. Next, we see that $\bar{G}_{\nu^{*}}(x)$ is continuous on $\left[0,U\right]$ and is a $C^{2} (\left(0,U)\right)$ function. In addition the boundary condition $\bar{G}_{\nu^{*}}(U)=1$ is satisfied as well. We have already shown that $-1 < d^{-} < 0$, and that $\frac{\partial}{\partial x}\bar{G}_{\nu^{*}}>0$, $\frac{\partial^{2}}{\partial x^{2}}\bar{G}_{\nu^{*}}<0$. Then, by substituting the maximizer into the wealth process we obtain for $t < \tau_{U}^{w^{*}}$
\begin{equation}
X^{*}_{\nu^{*}}(t)=\begin{cases} X(0)\exp\left\lbrace \left[\left(r+c^{net}\right)+\frac{\|\zeta\|^{2}}{1+d^{-}}\left(1-\frac{1}{2(1+d^{-})}\right)\right]t+\frac{1}{1+d^{-}}\zeta^{'} B(t)\right\rbrace & \mbox{if}\,\, \frac{1}{1+d^{-}}1_{N}^{'}\textit{\textbf{w}}_{o} < 1; \\
X(0)\exp\left\lbrace \beta_{\nu^{*}}t+\frac{1}{1+d^{-}}\zeta_{\nu^{*}}^{'}B(t) \right\rbrace & \mbox{if}\,\, \frac{1}{1+d^{-}}1_{N}^{'}\textit{\textbf{w}}_{o} \geq 1.
\label{444maximiserzvvv23} \end{cases}
\end{equation}

\noindent When the condition (\ref{condition}) holds and $\beta_{\nu^{*}} > 0$, the drift of $X^{*}_{\nu^{*}}(\cdot)$ is positive. Especially, by using the condition (\ref{condition}), we see for the first term in (\ref{444maximiserzvvv23}) that
\[r+c^{net} > - \frac{\|\zeta\|^{2}}{1+d^{-}}\left(1-\frac{1}{2(1+d^{-})}\right) > -\frac{\|\zeta\|^{2}}{2}. \]

\noindent Therefore, the markets are favourable and $\mathbb{P}\left(X^{*}_{\nu^{*}}(t) \geq U\right) \rightarrow 1$ as $t \rightarrow \infty$. This, in turn gives $\tau_{U}^{w^{*}} < \infty$ almost surely. It follows that
\begin{equation}
\int_{0}^{\tau^{w^{*}}_{U} \wedge \tau^{w^{*}}_{0}}\|\textit{\textbf{w}}^{*}(X^{*}_{\nu^{*}}(s))\|^{2}ds=\begin{cases} \frac{1}{(1+d^{-})^{2}}\|\textit{\textbf{w}}_{o}\|^{2} \tau^{w^{*}}_{U} < \infty & \mbox{if}\,\, \frac{1}{1+d^{-}}1_{N}^{'}\textit{\textbf{w}}_{o} < 1; \\ \frac{1}{K}\Sigma^{-1} \tau_{U}^{w^{*}}  < \infty & \mbox{if}\,\, \frac{1}{1+d^{-}}1_{N}^{'}\textit{\textbf{w}}_{o} \geq 1   \label{strategy1} \end{cases}
\end{equation}

\noindent almost surely. This gives $\textit{\textbf{w}}^{*} \in \mathcal{A}_{c}(x)$ for the maximization problem. Next, we introduce the function
\begin{equation}
M_{\nu^{*}}\left(t,X^{w}_{\nu^{*}}(t)\right)=e^{-\rho t}\bar{G}\left(X^{w}_{\nu^{*}}(t)\right). \label{ii3vpr1}
\end{equation}

\noindent Then, we define the stopping time
\begin{equation}
\tau_{n}^{w}=\tau_{U}^{w} \wedge \inf\{t > 0 \mid \int_{0}^{t} \|\textit{\textbf{w}}(X^{w}_{\nu^{*}}(s))\|^{2}ds = n\} \quad \mbox{for} \quad n \in \mathbb{N}, \nonumber
\end{equation}

\noindent and fix $x \in \left(0, U\right)$. We choose a strategy $\textit{\textbf{w}} \in \mathcal{A}_{c}(x)$ and write by It\^{o}'s formula
\begin{eqnarray}
M_{\nu^{*}}\left(\tau_{n}^{w},X^{w}_{\nu^{*}}(\tau_{n}^{w})\right)&=&\bar{G}_{\nu^{*}}\left(x\right)+\int_{0}^{\tau_{n}^{w}}e^{-\rho s}\{-\rho \bar{G}_{\nu^{*}}\left(X^{w}_{\nu^{*}}(s)\right)+\mathcal{L}^{w}\bar{G}_{\nu^{*}}\left(X^{w}_{\nu^{*}}(s)\right)\}ds \nonumber \\
& & \quad +\,\int_{0}^{\tau_{n}^{w}}X^{w}_{\nu^{*}}(s)\textit{\textbf{w}}^{'}(X^{w}_{\nu^{*}}(s))\sigma \frac{\partial \bar{G}_{\nu^{*}}}{\partial x}\left(X^{w}_{\nu^{*}}(s)\right)dB(s). \nonumber \\ \label{ii3vpr2}
\end{eqnarray} 

\noindent By using the value function $\bar{G}_{\nu^{*}}\left(X^{w}_{\nu^{*}}(s)\right)$ we can show that the stochastic integral above vanishes when its expectation is taken. That is, we have
\begin{eqnarray}
&&\mathbb{E}_{x}\left[\int_{0}^{\tau_{n}^{w}}\|X^{w}_{\nu^{*}}(s)\textit{\textbf{w}}^{'}(X^{w}_{\nu^{*}}(s))\sigma \frac{\partial \bar{G}_{\nu^{*}}}{\partial x}\left(X^{w}_{\nu^{*}}(s)\right)\|^{2}ds \right] \nonumber \\
&\leq &\Sigma(-d^{-})^{2}n < \infty . \label{ii3vpr3} 
\end{eqnarray}

\noindent On the other hand, because $\bar{G}_{\nu^{*}}(x)$ solves the HJB equation, we have $\forall \textit{\textbf{w}} \in \mathbb{R}^{N}$, and for each $s$ and $\mathbb{P}$-a.s., the inequality
\begin{equation}
-\rho \bar{G}_{\nu^{*}}\left(X^{w}_{\nu^{*}}(s)\right)+\mathcal{L}^{w}\bar{G}_{\nu^{*}}\left(X^{w}_{\nu^{*}}(s)\right) \leq 0. \label{ii3vpr5}
\end{equation}

\noindent As a result, we obtain the inequality
\begin{equation}
\bar{G}_{\nu^{*}}\left(x\right) \geq \mathbb{E}_{x}\left[M_{\nu^{*}}\left(\tau_{n}^{w}, X^{w}_{\nu^{*}}(\tau_{n}^{w})\right)\right]. \label{ii3vpr6}
\end{equation}

\noindent Because $\bar{G}_{\nu^{*}}(x) \in [0,1]$ and $\tau^{w}_{n} \rightarrow \tau^{w}_{U}$ as $n \rightarrow \infty$, from the dominated convergence theorem, we have as $n \rightarrow \infty$
\begin{equation}
\mathbb{E}_{x}\left[M_{\nu^{*}}(\tau_{n}^{w}, X^{w}_{\nu^{*}}(\tau_{n}^{w}))\right] \rightarrow \mathbb{E}_{x}\left[M_{\nu^{*}}(\tau^{w}_{U}, X^{w}_{\nu^{*}}(\tau^{w}_{U}))\right]. \nonumber
\end{equation}

\noindent Thus, the inequality in (\ref{ii3vpr6}) becomes
\begin{equation}
\bar{G}_{\nu^{*}}\left(x\right) \geq \mathbb{E}_{x}\left[M_{\nu^{*}}\left(\tau^{w}_{U}, X^{w}_{\nu^{*}}(\tau^{w}_{U})\right)\right]. \label{ii3vpr8}
\end{equation}

\noindent We have $\bar{G}_{\nu^{*}}(U)=1$. Therefore, we write from (\ref{ii3vpr8})
\begin{eqnarray}
\bar{G}_{\nu^{*}}(x) &\geq & \mathbb{E}_{x}\left[M_{\nu^{*}}\left(\tau^{w}_{U}, X^{w}_{\nu^{*}}(\tau^{w}_{U})\right)\right] \nonumber \\
&=&\mathbb{E}_{x}\left[e^{-\rho \tau_{U}^{w}}\right]. \label{ii3vpr9}
\end{eqnarray}

\noindent When we take the supremum over admissible strategies, we have
\[\bar{G}_{\nu^{*}}\left(x\right)\geq \sup_{\textit{\textbf{w}} \in \mathcal{A}_{c}}\mathbb{E}_{x}\left[e^{-\rho \tau_{U}^{w}}\right]=\bar{G}_{\nu^{*}}\left(x\right).\]

\noindent Therefore, $\textit{\textbf{w}}^{*}(x)$ is the optimal portfolio strategy, $X_{\nu^{*}}^{*}(t)$ is the optimal wealth process,  and $\bar{G}_{\nu^{*}}(x)$ is the optimal value function. In fact, when we choose the strategy $\textit{\textbf{w}}^{*}(X_{\nu^{*}}^{*}(t))$, we can see from (\ref{ii3vpr3}) that the stochastic integral vanishes, the inequality in (\ref{ii3vpr5}) becomes an equality making (\ref{ii3vpr6}) equal and showing the optimality of our results again.
\\

\noindent For the minimization problem we write $\textit{\underbar{G}}(x)=-\sup_{\textit{\textbf{w}} \in \mathcal{A}_{\nu}} \mathbb{E}_{x}\left[-e^{-\rho\tau_{L}^{w}}\right]$. Then, we apply the approach above to $\hat{G}(x)=-\textit{\underbar{G}}(x)$ for the proof of the results. However, notice that when the minimising strategy is substituted into the wealth process we obtain
\begin{equation}
X^{*}_{\nu^{*}}(t)=\begin{cases} X(0)\exp\left\lbrace \left[\left(r+c^{net}\right)+\frac{\|\zeta\|^{2}}{1+d^{+}}\left(1-\frac{1}{2(1+d^{+})}\right)\right]t+\frac{1}{1+d^{+}}\zeta^{'} B(t)\right\rbrace & \mbox{if}\,\, \frac{1}{1+d^{+}}1_{N}^{'}\textit{\textbf{w}}_{o} < 1; \\
X(0)\exp\left\lbrace \beta_{\nu^{*}}t+\frac{1}{1+d^{+}}\zeta_{\nu^{*}}^{'}B(t) \right\rbrace & \mbox{if}\,\, \frac{1}{1+d^{+}}1_{N}^{'}\textit{\textbf{w}}_{o} \geq 1.
\label{3pr15} \end{cases}
\end{equation}

\noindent This time, we are under the unfavourable markets. Therefore, we have $\beta_{\nu} < 0$ $\forall \nu \in \tilde{\mathcal{K}}$. From this, we see that the drift of the wealth process in the second case above is negative. For the first case, we have
\[r+c^{net} < -\frac{\|\zeta\|^{2}}{2} < - \frac{\|\zeta\|^{2}}{1+d^{+}}\left(1-\frac{1}{2(1+d^{+})}\right) < 0. \]

\noindent Therefore, $\mathbb{P}\left(X^{*}_{\nu^{*}}(t) > L\right) \rightarrow 0$ as $t \rightarrow \infty$. This gives $\tau^{w^{*}}_{L} < \infty$ almost surely. From this, we can show similarly that the minimising strategy $\textit{\textbf{w}}^{*}(x) \in \mathcal{A}_{c}(x)$ as well. The rest follows similarly.

\end{proof}

\section{Conclusion.} We provided examples concerning survival, growth, and goal reaching maximization under infinite time approach and borrowing constraints. In order to solve these problems, an auxiliary market is constructed and a dynamic programming approach is used. We can see in all problems that the optimal investment strategy is a constant proportional strategy that is independent of barrier levels and current wealth. Furthermore, the optimal investment strategy of the probability maximization and expected discounted reward maximization/minimization problems are proportional to log-optimal investment strategy. On the other hand, the survival time maximising strategy turned out to be the log-optimal investment strategy. Especially, when the market conditions are unfavourable, log-optimal strategy is the investment strategy that maximises survival time, and when the market conditions are favourable the log-optimal strategy is the investment strategy that minimises the time to reach a goal. When compared with the strategy of the probability maximization problem, we found that, under the unfavourable condition, the investor would take more risk to maximise the probability of hitting a target before hitting a lower boundary, compared to the risk she takes to maximise the expected time to hit the lower boundary. The situation is reversed when the conditions are favourable. The investor would take less risk to minimise the probability of ruin than the risk she takes to reach a higher target. Finally, when the markets are favourable there is no penalty to pay, and when unfavourable there is no reward to gain. In addition, under the favourable market conditions, a reward maximising investor would take more risk than she would with the log-optimal strategy. However, under the unfavourable markets a penalty minimising investor would take less risk than she would with the log-optimal strategy. 

\section*{Acknowledgments} For the paper, I used the results of my doctoral dissertation at Imperial College London. I would like to thank Mark H.A. Davis at Imperial College London for his supervision throughout my research. I also thank to Antoine Jacquier at Imperial College London for his helpful suggestions, and to Ege Yazgan at Istanbul Bilgi University for his support.

\nocite{Yener}

\bibliographystyle{elsart-harv}
\bibliography{MOR}

\appendix

\section{Appendix} The following is outlined from \citet{KaratzasShreve}. Notice that $X_{\nu}^{w}(t)\geq X^{w}(t)$ $\forall \nu \in \tilde{\mathcal{K}}$ Lebesgue-a.e. $t \in \left[0,\tau^{w}_{L} \wedge \tau_{U}^{w}\right]$. To see why this holds, let for $t \leq \tau^{w}_{L} \wedge \tau_{U}^{w}$
\begin{equation}
\chi(t)=\tilde{X}_{\nu}^{w}(t)-\tilde{X}^{w}(t), \nonumber 
\end{equation}

\noindent where \,$\tilde{}$\, over the portfolio processes denotes the discounted value. Then, we can write
\begin{eqnarray}
\chi(t)&=&\int_{0}^{t}\chi(s)\left(c^{net}+\delta(\nu)\right)ds+\int_{0}^{t}\chi(s)\textit{\textbf{w}}^{'}(s)\left(\mu-r1_{N}+\nu\right)ds  \nonumber\\
& & \quad +\,\int_{0}^{t}\chi(s)\textit{\textbf{w}}^{'}(s)\sigma dB(s)+\int_{0}^{t}\tilde{X}^{w}(s)\left(\delta(\nu)+\textit{\textbf{w}}^{'}(s)\nu\right)ds. \label{ieq12}
\end{eqnarray}

\noindent Next, we define the non-negative process
\begin{eqnarray}
M(t)&=&\exp\left\lbrace -\int_{0}^{t}\left(c^{net}+\delta(\nu)\right)ds-\int_{0}^{t}\textit{\textbf{w}}^{'}(s)\sigma dB(s)\right. \nonumber \\
&& \qquad \left. +\,\frac{1}{2}\int_{0}^{t}\|\sigma^{'}\textit{\textbf{w}}(s)\|^{2}ds -\int_{0}^{t}\textit{\textbf{w}}^{'}(s)\left(\mu -r1_{N}+\nu\right)ds \right\rbrace , \label{ieq13}
\end{eqnarray}

\noindent and compute the differential
\begin{eqnarray}
d\left(\chi(t)M(t)\right)=M(t)\tilde{X}^{w}\left(\delta(\nu)+\textit{\textbf{w}}^{'}(t)\nu\right)dt. \label{ieq14}
\end{eqnarray}

\noindent As a result, from the fact that $\textit{\textbf{w}}^{'}(t)\nu+\delta(\nu) \geq 0$ $\forall \nu \in \tilde{\mathcal{K}}$ Lebesgue-almost-every $t \in \left[0,\tau^{w}_{L} \wedge \tau_{U}^{w}\right]$ and that $\chi(0)=0$, the value in (\ref{ieq14}) is non-negative. Therefore, we have $X_{\nu}^{w}(t) \geq X^{w}(t)$. However, whenever $\textit{\textbf{w}}^{'}(t)\nu+\delta(\nu)=0$ for Lebesgue-almost-every $t \in \left[0,\tau^{w}_{L} \wedge \tau_{U}^{w}\right]$, $X_{\nu}^{w}(t) = X^{w}(t)$ almost surely. 

The above result implies that $\mathcal{A}_{c}(x) \subseteq \mathcal{A}_{\nu}(x)$. This gives $F_{c}(x) \leq F_{\nu}(x)$ $\forall \nu \in \tilde{\mathcal{K}}$. Therefore, when we find $\nu^{*} \in \tilde{\mathcal{K}}$ making $\textit{\textbf{w}}^{'}(t)\nu^{*}+\delta(\nu^{*})=0$ for Lebesgue-almost-every $t \in \left[0,\tau^{w}_{L} \wedge \tau_{U}^{w}\right]$, we have

\[F_{c}(x)=F_{\nu^{*}}(x)=\inf_{\nu \in \tilde{\mathcal{K}}}\Gamma_{\nu}(x),\]

\noindent and $\textit{\textbf{w}} \in \mathcal{K}$.

\begin{remark}
The above can also be applied to other problems we consider in this paper. Mainly, it can be applied similarly to $\textit{\underbar{F}}(x)$ and $\bar{G}(x)$. On the other hand, as we showed previously we can write  $\bar{F}(x)=-\sup_{\textit{\textbf{w}} \in \mathcal{A}_{\nu}} \mathbb{E}_{x}\left[-\tau_{U}^{w}\right]$ and $\textit{\underbar{G}}(x)=-\sup_{\textit{\textbf{w}} \in \mathcal{A}_{\nu}} \mathbb{E}_{x}\left[-e^{-\rho\tau_{L}^{w}}\right]$. Then, the above argument can be applied to the functions $\hat{F}(x)=-\bar{F}(x)$ and $\hat{G}(x)=-\textit{\underbar{G}}(x)$.
\end{remark}

\end{document}